\documentclass[12pt,american,english,intlimits]{article}
\usepackage[T1]{fontenc}
\usepackage[latin9]{inputenc}
\usepackage[a4paper]{geometry}
\usepackage{fancyhdr}
\usepackage{array}
\usepackage{longtable}
\usepackage{prettyref}
\usepackage{rotfloat}
\usepackage{textcomp}
\usepackage{url}
\usepackage{amsmath}
\usepackage{graphicx}
\usepackage{amssymb}
\usepackage[authoryear]{natbib}
\usepackage{nomencl}

\providecommand{\makenomenclature}{\makeglossary}
\makenomenclature

\makeatletter

\providecommand{\LyX}{L\kern-.1667em\lower.25em\hbox{Y}\kern-.125emX\@}
\providecommand{\tabularnewline}{\\}

\newrefformat{cap}{\hyperref[#1]{Fig.~\ref{#1}}}
\newrefformat{fig}{\hyperref[#1]{Fig.~\ref{#1}}}
\newrefformat{sec}{\hyperref[#1]{Sec.~\ref{#1}}}
\newrefformat{sub}{\hyperref[#1]{Sec.~\ref{#1}}}
\newrefformat{tab}{\hyperref[#1]{Table \ref{#1}}}
\newrefformat{eq}{\hyperref[#1]{eq.~(\ref{#1})}}

\@ifundefined{definecolor}
 {\usepackage{color}}{}

\definecolor{parametergray}{gray}{0.8}

\usepackage[OT1]{fontenc}

\renewenvironment{abstract}
{\noindent{\normalfont\large\textbf{Abstract}}%
\par\vspace{0.5\baselineskip}\noindent}
{\par}

\renewcommand{\@seccntformat}[1]{%
\csname the#1\endcsname\hspace{0.5em}}

\renewcommand{\section}{\@startsection
{section}%
{1}%
{0mm}%
{-\baselineskip}%
{0.5\baselineskip}%
{\normalfont\large\bfseries}}

\renewcommand{\subsection}{\@startsection
{subsection}%
{1}%
{0mm}%
{-\baselineskip}%
{0.5\baselineskip}%
{\normalfont\bfseries}}

\renewcommand{\subsubsection}{\@startsection
{subsubsection}%
{2}%
{1em}%
{-\baselineskip}%
{-\fontdimen2\font plus -\fontdimen3\font minus -\fontdimen4\font}%
{\normalfont\bfseries}}

\renewcommand{\@makecaption}[2]{%
{\parbox[t]{\linewidth}{%
\normalsize\renewcommand{\baselinestretch}{1.0}\normalsize 
\vspace{2mm}
\textbf{#1:} #2
}}}

\usepackage{babel}

\usepackage[%
colorlinks=true,linkcolor=blue,citecolor=blue,urlcolor=blue,filecolor=blue,%
pdffitwindow,%
pdftitle={The cell-type specific connectivity of the local cortical network explains prominent features of neuronal activity},
pdfauthor={Tobias C. Potjans},
pdfsubject={Potjans Layered Network Model},
pdfkeywords={Layer, Connectivity, Specificity, Simulation, Dynamics}
]{hyperref}

\usepackage{colortbl}

\usepackage{multirow}
\usepackage{rotate}

\makeatother

\usepackage{babel}

\begin{document}
\global\long\def\taum{\tau_{\text{m}}}
\global\long\def\taur{\tau_{\text{r}}}
\global\long\def\taua{\tau_{\alpha}}
\global\long\def\mm{\:\mathrm{mm}}
\global\long\def\mV{\:\mathrm{mV}}
\global\long\def\Hz{\:\mathrm{Hz}}
\global\long\def\ms{\:\mathrm{ms}}
\global\long\def\pA{\:\mathrm{pA}}
\global\long\def\pF{\:\mathrm{pF}}
\global\long\def\cm{C_{\mathrm{m}}}
\global\long\def\cp{C_{\mathrm{p}}}
\global\long\def\ca{C_{\mathrm{a}}}
\global\long\def\mum{\:\mu\mathrm{m}}
\global\long\def\dwp{\Delta w^{+}}
\global\long\def\dwm{\Delta w^{-}}
\global\long\def\s{\:\mathrm{s}}
\nomenclature{}{}\global\long\def\taufac{\tau_{\mathrm{fac}}}
\global\long\def\a{\mathrm{a}}
\global\long\def\p{\mathrm{p}}
\global\long\def\e{\mathrm{e}}
\global\long\def\i{\mathrm{i}}

\begin{titlepage}\thispagestyle{empty}\setcounter{page}{0}\pdfbookmark[1]{Title}{TitlePage}

\begin{center}
\textbf{\LARGE The cell-type specific connectivity of the local cortical
network explains prominent features of neuronal activity}
\par\end{center}{\LARGE \par}

\begin{center}
\textbf{\large Tobias C. Potjans$^{1,2,3}$, Markus Diesmann$^{1,2,4}$}
\par\end{center}{\large \par}

\vfill{}

\begin{flushleft}
\parbox[t]{17cm}{$^{1}$\mbox{Institute of Neuroscience and Medicine
(INM-6)}\\
\mbox{\hspace*{0.2cm}Computational and Systems Neuroscience}\\
\mbox{\hspace*{0.2cm}Research Center Juelich, 52425 Juelich, Germany}
\\
\mbox{$^{2}$Brain and Neural Systems Team}\\
\mbox{\hspace*{0.2cm}RIKEN Computational Science Research Program}\\
\mbox{\hspace*{0.2cm}Wako-shi, Saitama 351-0198, Japan} \\
\mbox{$^{3}$Faculty of Biology III}\\
\mbox{\hspace*{0.2cm}Albert-Ludwigs-University Freiburg}\\
\mbox{\hspace*{0.2cm}Schaenzlestrasse 1, 79104 Freiburg, Germany\hspace*{9cm}}
\mbox{$^{4}$RIKEN Brain Science Institute}\\
\mbox{\hspace*{0.2cm}Wako-shi, Saitama 351-0198, Japan}}
\par\end{flushleft}

\vspace{1cm}

\begin{abstract}\pdfbookmark[1]{Abstract}{AbstractPage}

In the past decade, the cell-type specific connectivity and activity
of local cortical networks have been characterized experimentally
to some detail. In parallel, modeling has been established as a tool
to relate network structure to activity dynamics. While the available
connectivity maps have been used in various computational studies,
prominent features of the simulated activity such as the spontaneous
firing rates do not match the experimental findings. Here, we show
that the inconsistency arises from the incompleteness of the connectivity
maps. Our comparison of the most comprehensive maps \citep{Thomson02_936,Binzegger04}
reveals their main discrepancies: the lateral sampling range and the
specific selection of target cells. Taking them into account, we compile
an integrated connectivity map and analyze the unified map by simulations
of a full scale model of the local layered cortical network. The simulated
spontaneous activity is asynchronous irregular and the cell-type specific
spontaneous firing rates are in agreement with \emph{in vivo} recordings
in awake animals, including the low rate of layer 2/3 excitatory cells.
Similarly, the activation patterns evoked by transient thalamic inputs
reproduce recent \emph{in vivo} measurements. The correspondence of
simulation results and experiments rests on the consideration of specific
target type selection and thereby on the integration of a large body
of the available connectivity data. The cell-type specific hierarchical
input structure and the combination of feed-forward and feedback connections
reveal how the interplay of excitation and inhibition shapes the spontaneous
and evoked activity of the local cortical network.

\end{abstract}

\end{titlepage}

\emph{}

\section*{Introduction\phantomsection\addcontentsline{toc}{section}{Introduction}\label{sec:Introduction}}

The local cortical network is considered a building block of brain
function and over the last century, the hypothesis that the interactions
of neurons within the microcircuit are governed by the cell-type specific
connectivity has been refined (see e.g. \citet{Douglas07_226,Douglas07_314}
for reviews). In the last decade, progress in methodology enabled
the compilation of comprehensive connectivity maps: \citet{Thomson02_936}
used electrophysiological recordings to estimate the connection probabilities
between various cell types in layers 2/3, 4 and 5 in slices of rat
and cat neocortex. Shortly thereafter, \citet{Binzegger04} applied
a modified version of Peters' rule \citep{Braitenberg98} to derive
the cell-type specific distribution of synapses from morphological
reconstructions of \emph{in vivo }labeled cells from area 17 of the
cat. These approaches entail partly contradicting results \citep{Thomson07_19},
but the compatibility of these maps and the influence of the different
methodologies have not been quantitatively assessed.

Due to recent advances in \emph{in vivo }electrophysiology and two-photon
optical imaging, the structural data can now be contrasted with observations
of the dynamics. Characteristic features are the cell-type specific
firing rates during ongoing activity in awake animals: low pyramidal
neuron firing rates below $1\Hz$ are reported in layer 2/3 (L2/3)
and highest rates in L5 \citep[e.g.][]{Greenberg08_749,deKock09_16446}.

Early network models already incorporate basic anatomy and electrophysiology
such as the separation of excitatory and inhibitory cell types and
a sparse, seemingly random connectivity \citep{Amit97,Vreeswijk96}.
The balance of excitation and inhibition explains the asynchronous
irregular (AI) spiking activity and the large membrane potential fluctuations
observed \emph{in vivo}. Models eventually incorporated multiple cell
types to capture layer-specific connections \citep[e.g. ][]{Hill05_1671,Traub05_2194}
and employ data based connectivity maps \citep{Haeusler07_149,Haeusler09_73,Heinzle07_9341,Izhikevich08_3593,Binzegger09_1071}.
However, no study to date reported cell-type specific firing rates
consistent with the experimental observations. A priori it is unclear
whether this is due to a misinterpretation of the raw connectivity
data or due to further model assumptions. Since the studies comprise
networks based on point-neuron as well as on multi-compartment neuron
models, the mismatch in fundamental characteristics like stationary
firing rates is unlikely to be caused by a lack in complexity of the
network elements but rather by the incompleteness of the connectivity
map.

In the present study we quantify the discrepancies between the electrophysiological
and anatomical connectivity maps to systematically compile an integrated
map. To resolve conflicts, we additionally incorporate insights from
photostimulation \citep{Dantzker00,Zarrinpar06_1751} and electron
microscopy \citep{McGuire84_3021} studies, reporting the specific
selection of interneurons by a subset of inter-layer projections.
We then check the consistency of structure and activity by means of
full scale simulations of the local cortical network.

\begin{figure}
\begin{centering}
\includegraphics[width=8.5cm]{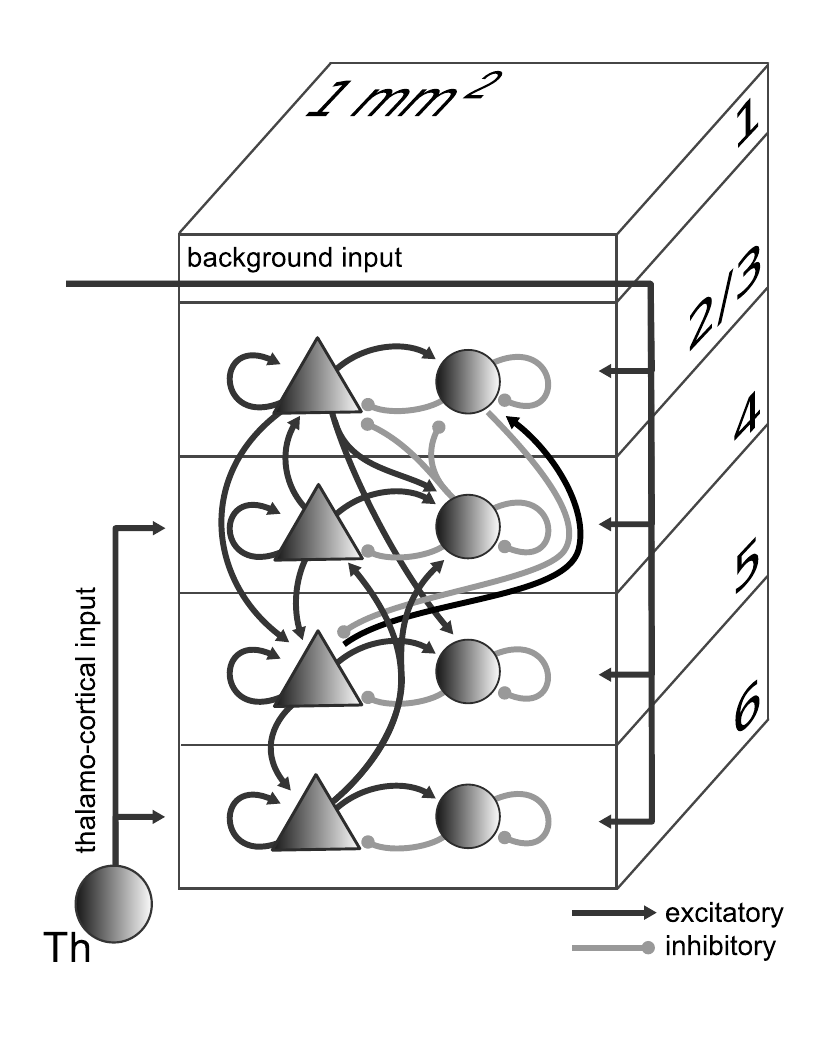}
\par\end{centering}

\caption{Model definition. Layers 2/3, 4, 5 and 6 are each represented by an
excitatory (triangles) and an inhibitory (circles) population of model
neurons. The number of neurons in a population is chosen according
to \citet{Binzegger04} based on the countings of \citet{Beaulieu83,Gabbott86_323}.
Input to the populations is represented by thalamo-cortical input
targeting layers 4 and 6 and background input to all populations.
Excitatory (black) and inhibitory (gray) connections with connection
probabilities $>0.04$ are shown. The model size corresponds to the
cortical network below 1 $\mathrm{mm}^{2}$ surface.\label{fig:Model-definition}}

\end{figure}

\section*{Materials and Methods \phantomsection\addcontentsline{toc}{section}{Materials and Methods} \label{sec:Materials-and-Methods}}

The network model (\prettyref{fig:Model-definition}) represents four
layers of cortex, L2/3, L4, L5 and L6, each consisting of two populations
of excitatory (e) and inhibitory (i) neurons. Throughout the article,
we use the term \emph{connection} with reference to populations, defined
by the pre- and postsynaptic layers and neuron types. The term \emph{projection}
is used for the two connections of a single presynaptic population
to both populations of a target layer. A connectivity map is defined
by the sixty-four connection probabilities between the eight considered
cell types.

\subsection*{Connectivity data\phantomsection\addcontentsline{toc}{subsection}{Connectivity data} \label{sub:Connectivity-data}}

For the anatomical map (a), \citet{Binzegger04} provide the relative
number of synapses participating in a connection and the total absolute
number of synapses, depending on pre- and postsynaptic type, of area
17 (supplementary Table 6). The product of these measures gives the
absolute number of synapses $K$ for any connection. To calculate
the corresponding connection probabilities $C_{\mathrm{a}}$ we assume
that the synapses are randomly distributed allowing multiple contacts
between any two neurons. With $N^{\mathrm{pre(post)}}$ being the
number of neurons in the presynaptic (postsynaptic) population: 

\begin{eqnarray}
C_{\mathrm{a}} & = & 1-\left(1-\frac{1}{N^{\mathrm{pre}}N^{\mathrm{post}}}\right)^{K}.\label{eq:connection_prob}\end{eqnarray}
The often used expression\begin{eqnarray}
C_{\a} & = & K/N^{\mathrm{pre}}N^{\mathrm{post}}\label{eq:approx_conn_prob}\end{eqnarray}
is the corresponding first-order Taylor series approximation and valid
for small $K/(N^{\mathrm{pre}}N^{\mathrm{post}})$ (see \emph{Supplemental
Material}). The original published data constitute our raw\emph{
}connectivity map (\citet{Binzegger04}, their figure 12). Consistent
with other modeling work \citep{Izhikevich08_3593}, we construct
an improved ({}``modified'') anatomical map by assigning the unassigned
symmetric (inhibitory) synapses, originating from a potential underestimation
of interneuronal connectivity, to within-layer projections originating
from local interneurons \citep{Binzegger04}. The derived connection
probabilities are inversely proportional to the considered surface
area $\pi r^{2}$ (\prettyref{fig:connmatrixproperties}B). This can
be easily understood when considering the approximation \prettyref{eq:approx_conn_prob}:
the numbers of neurons $N$ and synapses $K$ increase linear with
the surface area and therefore $C_{\mathrm{a}}\propto\pi r^{2}/(\pi r^{2})^{2}=1/\pi r^{2}$.
The product of the connection probability and the surface area is
constant for large areas (\prettyref{fig:connmatrixproperties}B).
Hence, we use throughout this article the area-corrected connection
probability $\tilde{C}_{\mathrm{a}}=\lim_{r\to\infty}C_{\mathrm{a}}\pi r^{2}$
for all numerical values of the anatomical map.

The physiological hit rate estimates (\citet{Thomson02_936}, their
Table 1) provide the physiological map (p). We combine multiple independently
measured hit rates for the same connection by a weighted sum

\begin{eqnarray}
C_{\mathrm{p}} & =\sum_{i}R_{i}Q_{i}/\sum_{j}Q_{j} & ,\label{eq:conn_prob_paired}\end{eqnarray}
where $R_{i}$ and $Q_{i}$ are the hit rate and the number of tested
pairs in the $i$th experiment, respectively. In accordance with \citet{Haeusler07_149},
we set the probabilities of the L2/3i to L5e and of the L4i to L2/3i
connections to $0.2$. While these data constitute the raw map, we
incorporate additional hit rate estimates (\prettyref{tab:Physiological-data})
to create an improved (modified) physiological connectivity map. The
numerical values of all connectivity maps are listed in supplementary
Table 6.

We classify all connections into two main groups: recurrent intra-layer
or {}``within-layer'' connections and connections between different
layers or {}``inter-layer'' connections.

\subsection*{Lateral connectivity model\phantomsection\addcontentsline{toc}{subsection}{Lateral connectivity model} \label{sub:Lateral-connectivity-model}}

We use a Gaussian model for describing the lateral connection probability
profile \begin{eqnarray}
C(r) & = & C_{0}\exp\left(\frac{-r^{2}}{2\sigma^{2}}\right).\label{eq:Gauss}\end{eqnarray}
with $r$ being the lateral distance and $C_{0}$ and $\sigma$ specifying
the peak connection probability (zero lateral distance) and the lateral
spread of connections, respectively. Thereby, we assume that the underlying
lateral connectivity is the same for both connectivity maps, i.e.
that $C_{0}$ and $\sigma$ are universal. Furthermore, we assume
that the experimental data correspond to a random sampling of connections
within a cylinder of a fixed sampling radius $r_{\mathrm{a}}>r_{\mathrm{p}}$
and corresponding mean connection probabilities $C_{\mathrm{a}/\p}=1/(\pi r_{\a/\p}^{2})\int_{0}^{r_{\mathrm{a}/\p}}\int_{0}^{2\pi}C(r)rdrd\varphi=2\pi C_{0}\sigma^{2}/(\pi r_{\a/\p}^{2})[1-\exp(-r_{\a/\p}^{2}/2\sigma^{2})]$.
These expressions for the mean connection probabilities of the two
connectivity maps allow us to determine the two unknown parameters
$C_{0}$ and $\sigma$. We obtain in closed form (see \emph{Supplemental
Material})\begin{eqnarray}
\sigma & = & r_{\mathrm{p}}\left[-2\ln\left(1-\frac{\pi r_{\mathrm{p}}^{2}C_{\mathrm{p}}}{\tilde{C}_{\mathrm{a}}}\right)\right]^{-1/2}\label{eq:sigma}\end{eqnarray}
and

\begin{eqnarray}
C_{0} & = & \frac{\tilde{C}_{\mathrm{a}}}{2\pi\sigma^{2}}.\label{eq:c0}\end{eqnarray}
In principle, this approach may be applied to an individual connection.
However, the sampling radius is the same for all connections of a
map. Therefore we determine $\sigma$ and $C_{0}$ only for the global
mean connection probabilities of the two maps which provides robustness
against uncertainties in the probability estimate of a particular
connection. 

The simulations use a laterally uniform connectivity profile, i.e.
the connectivity between two neurons is exclusively determined by
the cell-types and not their location in space. The mean connection
probability of the model $C_{\mathrm{m}}$ depends on the size of
the network (e.g. the surface area $\pi r_{\mathrm{m}}^{2}$) and
the parameters of the lateral model, but by applying eqs. \eqref{eq:sigma}
and \eqref{eq:c0} it can also be expressed in terms of the experimentally
accessible parameters $C_{\mathrm{p}}$, $r_{\mathrm{p}}$ and $\tilde{C}_{\mathrm{a}}$
\begin{eqnarray}
C_{\mathrm{m}} & = & \frac{1}{\pi r_{\mathrm{m}}^{2}}\int_{0}^{r_{\mathrm{m}}}\int_{0}^{2\pi}C(r)rdrd\varphi\nonumber \\
 & = & \frac{2}{r_{\mathrm{m}}^{2}}C_{0}\sigma^{2}[1-\exp(-r_{\mathrm{m}}^{2}/(2\sigma^{2}))]\nonumber \\
 & = & \frac{\tilde{C}_{\mathrm{a}}}{\pi r_{\mathrm{m}}^{2}}\left[1-(1-\pi r_{\mathrm{p}}^{2}\bar{C}_{\mathrm{p}}/\bar{\tilde{C}}_{\mathrm{a}})^{r_{\mathrm{m}}^{2}/r_{\mathrm{p}}^{2}}\right].\label{eq:cm3}\end{eqnarray}
where $\bar{C}_{\mathrm{p}}$, and $\bar{\tilde{C}}_{\mathrm{a}}$
specify the global means. To arrive at the individual connection probabilities
at a given model size it is sufficient to multiply a connectivity
map by the ratio of $C_{\mathrm{m}}$ and the global mean of the map
$\bar{C}_{\mathrm{a/p}}$.

\begin{longtable}{l>{\raggedright}p{1.2cm}>{\raggedright}p{1.2cm}l}
\caption{Modified physiological connectivity map. The map incorporates hit
rate estimates of several studies. The selection of studies is based
on the comprehensive review by \citet{Thomson07_19} including all
data where hit rate and the number of tested pairs can be extracted
such that \prettyref{eq:conn_prob_paired} is applicable. The table
lists, from left to right, connection specifier, number of existing
connections (product of hit rate and number of tested pairs), number
of tested pairs and the publication from which the data are extracted.
A star ({*}) indicates that the number of tested pairs is not explicitly
given, but estimated from the stated accuracy of connection probability.
Furthermore, we use the following additional data on within-layer
connections that is not reported separately. \citet{Thomson96_81,Thomson07_19}:
i to e in L2/3, L4 and L5 $21$ connected of $93$ tested pairs; \citet{Ali07_149}:
i to e in L2/3, L4 and L5 $30$ connected of $90$ tested pairs, e
to i in L2/3, L4 and L5 $21$ connected of $48$ tested pairs; i to
e in L2/3 and L4 $9$ connected of $69$ tested pairs and e to i in
L2/3 and L4 $21$ connected of $147$ tested pairs. The reported numbers
of connected and tested pairs are uniformly distributed to the reported
layers.\label{tab:Physiological-data}}
\tabularnewline
connection & existing & tested & publication\tabularnewline
\hline
\endfirsthead
connection & existing & tested & publication\tabularnewline
\hline
\hline
\endhead
\hline 
L2/3e$\to$L2/3e & 65 & 247 & \citet{Thomson02_936} (rat)\tabularnewline
 & 8 & 81 & \citet{Thomson02_936} (cat)\tabularnewline
 & 8 & 32 & \citet{Bannister07_2190} (rat)\tabularnewline
 & 3 & 36 & \citet{Bannister07_2190} (cat)\tabularnewline
 & 48 & 549 & \citet{Mason91}\tabularnewline
 & 32 & 305 & \citet{Kapfer07_743}\tabularnewline
 & 22 & 112 & \citet{Yoshimura05_868}\tabularnewline
 & 63 & 760 & \citet{Holmgren03_139}\tabularnewline
 & 24 & 110 & \citet{Ren07_758}\tabularnewline
\hline 
L2/3e$\to$L2/3i & 6 & 25 & \citet{Thomson02_936} (cat)\tabularnewline
 & 22 & 107 & \citet{Thomson02_936} (rat)\tabularnewline
 & 151 & 243 & \citet{Holmgren03_139}\tabularnewline
 & 19 & 40 & \citet{Kapfer07_743} (FS)\tabularnewline
 & 29 & 100 & \citet{Kapfer07_743}(SOM)\tabularnewline
\hline 
L2/3i$\to$L2/3e & 7 & 25 & \citet{Thomson02_936} (cat)\tabularnewline
 & 17 & 107 & \citet{Thomson02_936} (rat)\tabularnewline
 & 136 & 243 & \citet{Holmgren03_139}\tabularnewline
 & 26 & 39 & \citet{Kapfer07_743} (FS)\tabularnewline
 & 19 & 39 & \citet{Kapfer07_743}(SOM)\tabularnewline
\hline 
L2/3i$\to$L2/3i & 2 & 2 & \citet{Thomson02_936} (cat)\tabularnewline
 & 2 & 8 & \citet{Thomson02_936} (rat)\tabularnewline
\hline
\hline 
L4e$\to$L4e & 10 & 139 & \citet{Bannister07_2190} (cat)\tabularnewline
 & 22 & 528 & \citet{Bannister07_2190} (rat)\tabularnewline
 & 4 & 23 & \citet{Thomson02_936}\tabularnewline
 & 131 & 655 & \citet{Feldmeyer05_3423}\tabularnewline
 & 11 & 89 & \citet{Beierlein03_2987}\tabularnewline
 & 25 & 234 & \citet{Maffei04_1353}\tabularnewline
\hline 
L4e$\to$L4i & 8 & 42 & \citet{Thomson02_936}\tabularnewline
 & 3 & 21 & \citet{Ali07_149} (nonFS)\tabularnewline
 & 11 & 154 & \citet{Ali07_149} (FS)\tabularnewline
 & 74 & 172 & \citet{Beierlein03_2987} (FS)\tabularnewline
 & 36 & 63 & \citet{Beierlein03_2987} (LTS)\tabularnewline
\hline 
L4i$\to$L4e & 10 & 64 & \citet{Ali07_149} (nonFS)\tabularnewline
 & 10 & 40 & \citet{Ali07_149} (FS)\tabularnewline
 & 83 & 190 & \citet{Beierlein03_2987} (FS)\tabularnewline
 & 26 & 74 & \citet{Beierlein03_2987} (LTS)\tabularnewline
 & 4 & 42 & \citet{Thomson02_936}\tabularnewline
\hline 
L4i$\to$L4i & 3 & 6 & \citet{Thomson02_936}\tabularnewline
\hline
L5e$\to$L5e & 15 & 163 & \citet{Thomson02_936}\tabularnewline
 & 50 & 500 & \citet{Markram97b}\tabularnewline
 & 218 & 1655 & \citet{LeBe2006_13214}\tabularnewline
 & 29 & 206 & \citet{LeBe2006_13214}\tabularnewline
 & 148 & 1233 & \citet{Wang06_534}\tabularnewline
 & 26 & 260 & \citet{Wang06_534}\tabularnewline
 & 173 & 1450 & \citet{Silberberg07_735}\tabularnewline
\hline 
L5e$\to$L5i & 19 & 190 & \citet{Thomson97_131,Thomson97_510}{*}\tabularnewline
 & 6 & 79 & \citet{Thomson95_727}\tabularnewline
 & 7 & 73 & \citet{Thomson02_936}\tabularnewline
\hline 
L5i$\to$L5e & 9 & 73 & \citet{Thomson02_936}\tabularnewline
\hline 
L5i$\to$L5i & 3 & 5 & \citet{Thomson02_936}\tabularnewline
\hline
\hline 
L6e$\to$L6e & 56 & 1512 & \citet{Mercer05_1485}\tabularnewline
 & 4 & 204 & \citet{Beierlein02_1924}\tabularnewline
\hline 
L6e$\to$L6i & 8 & 38 & \citet{West06_200} (cat)\tabularnewline
 & 5 & 21 & \citet{West06_200} (rat)\tabularnewline
\hline
\hline 
L2/3e$\to$L4e & 0 & 25 & \citet{Thomson02_936} (rat)\tabularnewline
 & 0 & 70 & \citet{Thomson02_936} (cat)\tabularnewline
\hline 
L2/3e$\to$L4i & 1 & 12 & \citet{Thomson02_936} (rat)\tabularnewline
 & 7 & 37 & \citet{Thomson02_936} (cat)\tabularnewline
\hline 
L2/3i$\to$L4e & 0 & 29 & \citet{Thomson02_936} (rat)\tabularnewline
 & 0 & 10 & \citet{Thomson02_936} (cat)\tabularnewline
\hline 
L2/3e$\to$L5e & 2 & 2 & \citet{Thomson02_936} (cat)\tabularnewline
 & 16 & 29 & \citet{Thomson02_936} (rat)\tabularnewline
 & 25 & 259 & \citet{Thomson98_669}\tabularnewline
 & 247 & 1324 & \citet{Kampa06_1472}\tabularnewline
\hline
\hline 
L4e$\to$L2/3e & 10 & 50 & \citet{Yoshimura05_868}\tabularnewline
 & 7 & 25 & \citet{Thomson02_936} (rat)\tabularnewline
 & 7 & 70 & \citet{Thomson02_936} (cat)\tabularnewline
 & 64 & 640 & \citet{Feldmeyer02_803,Feldmeyer05_3423}\tabularnewline
\hline 
L4e$\to$L2/3i & 1 & 10 & \citet{Thomson02_936} (rat)\tabularnewline
 & 3 & 31 & \citet{Thomson02_936} (cat)\tabularnewline
\hline 
L4i$\to$L2/3e & 6 & 12 & \citet{Thomson02_936} (rat)\tabularnewline
 & 10 & 37 & \citet{Thomson02_936} (cat)\tabularnewline
\hline
\hline 
L4e$\to$L5e & 12 & 86 & \citet{Feldmeyer05_3423}{*}\tabularnewline
\hline
\hline 
L5e$\to$L2/3e & 1 & 29 & \citet{Thomson02_936}\tabularnewline
 & 3 & 259 & \citet{Thomson98_669}\tabularnewline
\end{longtable}

\subsection*{Connectivity data analysis\phantomsection\addcontentsline{toc}{subsection}{Connectivity data analysis} }

To compare the connectivity maps we define \textit{\emph{recurrence
strength}}\emph{ }as the ratio of the mean within-layer and the mean
inter-layer connection probabilities and \textit{\emph{loop strength}}\emph{
}as the ratio of the mean connection probability of the cortical feed-forward
loop (\citet{Gilbert83_217}, L4 to L2/3 to L5 to L6 and back to L4)
and the mean connection probability of all other inter-layer connections.
For a fair comparison we base recurrence and loop strength only on
connections for which estimates are available in both data sets. Therefore,
layers 2/3, 4 and 5 are included but not layer 6. 

A measure with higher resolution is the \textit{\emph{scaling factor}}\emph{
}$\zeta$ which compares the connection probabilities of individual
connections provided both connectivity maps assign non-zero probabilities.\begin{eqnarray}
\zeta & = & \frac{\max(C_{\mathrm{a}}/\bar{C}_{\mathrm{a}},C_{\mathrm{p}}/\bar{C}_{\mathrm{p}})}{\min(C_{\mathrm{a}}/\bar{C}_{\mathrm{a}},C_{\mathrm{p}}/\bar{C}_{\mathrm{p}})}\label{eq:scalingfactor}\end{eqnarray}
is independent of model size because $\cm$ cancels from the expression
(see \emph{Supplemental Material}).

Furthermore, to quantify the specificity of connections we introduce
the \textit{\emph{target specificity}}\begin{eqnarray}
T & = & \frac{C^{\mathrm{post=e}}-C^{\mathrm{post=i}}}{C^{\mathrm{post=e}}+C^{\mathrm{post=i}}}.\label{eq:targetspecificity}\end{eqnarray}
as the normalized difference of the connection probabilities constituting
a projection.

\subsection*{Consistent modifications of target specificity\phantomsection\addcontentsline{toc}{subsection}{Introducing target specificity} \label{sub:Render-target-specific}}

In order to construct a consistent integrated connectivity map, it
is necessary to modify the target specificity of certain projections
(see\emph{ Results}), i.e. connection probabilities are modified to
meet a given target specificity value. However, we constrain these
modifications by demanding consistency with the underlying experimental
data. For the anatomical data, this underlying measure is the number
of synapses participating in a projection. For the physiological data,
it is the measured connection probability of one of the two connections
forming the projection (typically the second connection has not been
quantified).

Modifying the connection probabilities while conserving the total
number of synapses of a projection requires a redistribution of synapses
across the target neurons (see supplementary Fig. 13). To that end,
we determine the fraction of synapses targeting excitatory neurons
$\Delta$ as a function of the requested target specificity constrained
by the total number of synapses and the sizes of the presynaptic and
the two postsynaptic populations. The main complication is that target
specificity is defined in terms of connection probabilities, see \prettyref{eq:targetspecificity},
and the relation of connection probability and the number of synapses
is non-linear, see \prettyref{eq:connection_prob} (for the exact
expression see \emph{Supplemental Material}). Nevertheless, in the
first-order Taylor series expansion of $C$, \prettyref{eq:approx_conn_prob},
the relation is linear and substituting $C^{\mathrm{post=e}}=\Delta K/(N^{\mathrm{pre}}N^{\mathrm{post=e}})$
and $C^{\mathrm{post=i}}=(1-\Delta)K/(N^{\mathrm{pre}}N^{\mathrm{post=i}})$
in \prettyref{eq:targetspecificity} we find

\begin{eqnarray*}
\Delta & = & \frac{(1+T)N^{\mathrm{post=e}}}{(1-T)N^{\mathrm{post=i}}+(1+T)N^{\mathrm{post=e}}}.\end{eqnarray*}

The modifications of the physiological data are straightforward because
in all cases considered here only one connection probability is experimentally
given so that we can estimate the unknown value based on the definition
\prettyref{eq:targetspecificity}: \begin{eqnarray*}
C^{\mathrm{post=i(e)}} & = & \left(\frac{1-T}{1+T}\right)^{+(-)1}C^{\mathrm{post=e(i)}}.\end{eqnarray*}

\subsection*{Compilation of the integrated connectivity map}

We compile the integrated connectivity map algorithmically. The procedure
requires as input the anatomical and the physiological connectivity
maps as well as the model size and information about desired modifications
of the target specificities. The procedure automatically estimates
the parameters of the Gaussian model and, based on these, the mean
model connectivity. Subsequently the procedure separately modifies
the target specificities of both scaled maps as instructed and finally
averages the two maps (see also \emph{Supplemental Material}).

\subsection*{Layer-specific external input\phantomsection\addcontentsline{toc}{subsection}{Layer-specific external input} }

We distinguish three types of layer-specific external inputs: thalamic
afferents as reconstructed by \citet{Binzegger04}, {}``gray-matter''
external inputs, i.e. intrinsic non-local inputs entering the local
network through the gray matter, which we estimate based on the properties
of axonal clusters \citep{Binzegger07_12242}, and {}``white-matter''
external inputs, which include all inputs not covered by the previous
types. The number of the latter is estimated based on the difference
of the number of synapses the different cell types receive according
to \citet{Binzegger04} and the total synapse count in a given layer
\citep{Beaulieu85}.

The thalamic afferents are included in the anatomical connectivity
map \citep{Binzegger04}. We extract the gray-matter inputs from the
information on bouton distributions in 3-dimensional space described
in \citet{Binzegger07_12242}. The authors find that boutons of all
cell types form multiple clusters and the article provides the lateral
distance of cluster centers and the corresponding somata. We interpret
a cluster to be non-local if the lateral distance to the soma is larger
than $\approx0.56\mm$ (corresponding to a local network surface area
of $1\mm^{2}$). By additionally using data on the relative sizes
of different cluster types, we estimate the proportion of intrinsic
gray-matter connections that originate outside of the local network.
Thereupon, we use the estimated proportion of gray-matter inputs and
the number of local connections in our network to calculate the absolute
number of gray-matter inputs. In this way we construct an estimate
of the gray-matter external inputs that is consistent with both, the
axonal structure in \citet{Binzegger07_12242} and the structure of
our model. The detailed procedure is described in the \emph{Supplemental
Material}.

The white-matter inputs are estimated based on the comparison of the
absolute number of synapses obtained in \citet{Binzegger04} which
only contains the contributions from local, thalamic and gray-matter
synapses, with those in \citet{Beaulieu85} containing all synapses.
The difference has been termed the {}``dark matter'' of cortex \citep{Binzegger04}
and, in case of the excitatory synapses, is usually interpreted as
white-matter external inputs. The explicit numbers are published in
\citeauthor{Izhikevich08_3593} \citeyearpar[their figure 9 of the supplemental material]{Izhikevich08_3593}
at subcellular resolution. As our model is based on point neurons,
we sum over all contributions to a given cell type and average across
the cell types that are collapsed to a single population in our model.
Thereby, our estimates take neuronal morphology into account. The
resulting counts for the three external input types as well as the
total number of external inputs to the excitatory populations are
given in supplementary Table 7. Since long-range projections target
excitatory as well as inhibitory neurons \citep{Johnson96_383,Gonchar03_10904},
we choose target specificity values for external inputs to be comparable
to recurrent connections, resulting in similar total numbers of external
inputs to inhibitory neurons (\prettyref{tab:Simulation-parameters}).

The cell-type specific external inputs are used as a reference parametrization
of the model. To comprehensively assess the dependence of the activity
on external inputs, we conduct simulations in which we choose the
external inputs to a population randomly with uniform distribution:
The inputs to the excitatory populations are constrained by the values
from the reference and the layer-independent parametrization (compare
supplementary Fig. 18). Those to inhibitory populations are chosen
such that the target specificity of the external input to a given
layer is between $0$ and $0.1$ (to L2/3, L4 and L5) and to L6, due
to the high number of external inputs to excitatory cells in this
layer, between $0$ and $0.2$. 

\begin{table}
\begin{tabular}{|@{\hspace*{1mm}}p{3cm}@{}|@{\hspace*{1mm}}p{12.2cm}|}
\hline 
\multicolumn{2}{|>{\color{white}\columncolor{black}}c|}{\textbf{A: Model summary}}\tabularnewline
\hline
\textbf{Populations} & nine; eight cortical populations and one thalamic population\tabularnewline
\hline 
\textbf{Topology} & ---\tabularnewline
\hline 
\textbf{Connectivity} & random connections\tabularnewline
\hline 
\textbf{Neuron model } & cortex: leaky integrate-and-fire, fixed voltage threshold, fixed absolute
refractory period (voltage clamp), thalamus: fixed-rate Poisson\tabularnewline
\hline 
\textbf{Synapse model} & exponential-shaped postsynaptic currents\tabularnewline
\hline 
\textbf{Plasticity} & ---\tabularnewline
\hline 
\textbf{Input} & cortex: independent fixed-rate Poisson spike trains \tabularnewline
\hline 
\textbf{Measurements} & spike activity, membrane potentials\tabularnewline
\hline
\end{tabular}

\begin{tabular}{|@{\hspace*{1mm}}p{3.cm}@{}|@{\hspace*{1mm}}p{2.8cm}@{}|@{\hspace*{1mm}}p{4.8cm}@{}|@{\hspace*{1mm}}p{4.4cm}|}
\hline 
\multicolumn{4}{|>{\color{white}\columncolor{black}}c|}{\textbf{B: Populations}}\tabularnewline
\hline 
\textbf{Type} & \textbf{Elements} & \textbf{Number of populations} & \textbf{Population size}\tabularnewline
\hline 
Cortical network & iaf neurons & eight, two per layer & $N$ (type-specific)\tabularnewline
\hline 
Th & Poisson & one & $N_{\mathrm{th}}$\tabularnewline
\hline
\end{tabular}

\begin{tabular}{|@{\hspace*{1mm}}p{3cm}@{}|@{\hspace*{1mm}}p{12.2cm}|}
\hline 
\multicolumn{2}{|>{\color{white}\columncolor{black}}c|}{\textbf{C: Connectivity}}\tabularnewline
\hline 
\textbf{Type} & random connections with independently chosen pre- and postsynaptic
neurons; see \prettyref{tab:Simulation-parameters} for probabilities\tabularnewline
\hline 
\textbf{Weights} & fixed, drawn from Gaussian distribution\tabularnewline
\hline 
\textbf{Delays} & fixed, drawn from Gaussian distribution multiples of sim. stepsize\tabularnewline
\hline
\end{tabular}

\begin{tabular}{|@{\hspace*{1mm}}p{3cm}@{}|@{\hspace*{1mm}}p{12.2cm}|}
\hline 
\multicolumn{2}{|>{\color{white}\columncolor{black}}c|}{\textbf{D: Neuron and synapse model}}\tabularnewline
\hline 
\textbf{Name} & iaf neuron\tabularnewline
\hline 
\textbf{Type} & leaky integrate-and-fire, exponential-shaped synaptic current inputs\tabularnewline
\hline 
\textbf{Subthreshold dynamics} & $\frac{dV}{dt}=-\frac{V}{\tau_{\mathrm{m}}}+\frac{I(t)}{C_{\mathrm{m}}}$\hspace*{0.1cm}
if $\mathrm{\left(t>t^{*}+\tau_{ref}\right)}$\newline $V(t)=V_{\mathrm{reset}}$\hspace*{1cm}
else 

$I_{\mathrm{syn}}(t)=w\, e^{-t/\tau_{\mathrm{syn}}}$\tabularnewline
\hline 
\textbf{Spiking} & If $V(t-)<\theta\wedge V(t+)\geq\theta$\newline1. set $t^{*}=t$,
2. emit spike with time stamp $t^{*}$\tabularnewline
\hline
\end{tabular}

\begin{tabular}{|@{\hspace*{1mm}}p{3cm}@{}|@{\hspace*{1mm}}p{3cm}@{}|@{\hspace*{1mm}}p{9.1cm}|}
\hline 
\multicolumn{3}{|>{\color{white}\columncolor{black}}c|}{\textbf{E: Input}}\tabularnewline
\hline 
\textbf{Type} & \textbf{Target} & \textbf{Description}\tabularnewline
\hline 
Background & iaf neurons & independent Poisson spikes (see \prettyref{tab:Simulation-parameters})\tabularnewline
\hline
\end{tabular}

\selectlanguage{american}%
\begin{tabular}{|l||c|}
\hline 
\multicolumn{2}{|>{\color{white}\columncolor{black}}c|}{\selectlanguage{english}%
\textbf{F: Measurements}\selectlanguage{american}
}\tabularnewline
\hline 
\multicolumn{2}{|p{15.2cm}|}{\selectlanguage{english}%
Spiking activity and membrane potentials from a subset of neurons
in every population\selectlanguage{american}
}\tabularnewline
\hline
\end{tabular}

\selectlanguage{english}%
\caption{Model description after \citet{Nordlie-2009_e1000456}. \label{tab:Model-description}}

\end{table}

\begin{table}
\begin{tabular}{|@{\hspace*{1mm}}p{3.5cm}@{}|@{\hspace*{1mm}}p{1.1cm}@{}|@{\hspace*{1mm}}p{1.1cm}@{}|@{\hspace*{1mm}}p{1.1cm}@{}|@{\hspace*{1mm}}p{1.1cm}|@{\hspace*{1mm}}p{1.1cm}|@{\hspace*{1mm}}p{1.1cm}|@{\hspace*{1mm}}p{1.1cm}|@{\hspace*{1mm}}p{1.1cm}|@{\hspace*{1mm}}p{0.7cm}|}
\hline 
\multicolumn{10}{|>{\columncolor{parametergray}}c|}{\textbf{B+E: Populations and Inputs}}\tabularnewline
\hline 
\textbf{Name} & L2/3e & L2/3i & L4e & L4i & L5e & L5i & L6e & L6i & Th\tabularnewline
\hline 
Population size $N$ & $20683$ & $5834$ & $21915$ & $5479$ & $4850$ & $1065$ & $14395$ & $2948$ & $902$\tabularnewline
\hline 
External inputs $k_{\mathrm{ext}}$ & $1600$ & $1500$ & $2100$ & $1900$ & $2000$ & $1900$ & $2900$ & $2100$ & ---\tabularnewline
\hline 
Background rate $\nu_{\mathrm{bg}}$ & \multicolumn{9}{@{\hspace*{1mm}}p{9.9cm}@{}|}{$8$ Hz}\tabularnewline
\hline
\end{tabular}

\begin{tabular}{|@{\hspace*{1mm}}p{1.4cm}@{}|@{\hspace*{1mm}}p{1.4cm}@{}|@{\hspace*{1mm}}p{1.1cm}@{}|@{\hspace*{1mm}}p{1.1cm}@{}|@{\hspace*{1mm}}p{1.1cm}@{}|@{\hspace*{1mm}}p{1.1cm}|@{\hspace*{1mm}}p{1.1cm}|@{\hspace*{1mm}}p{1.1cm}|@{\hspace*{1mm}}p{1.1cm}|@{\hspace*{1mm}}p{1.1cm}|@{\hspace*{1mm}}p{1.1cm}|}
\hline 
\multicolumn{11}{|>{\columncolor{parametergray}}c|}{\textbf{C: Connectivity}}\tabularnewline
\hline 
\multicolumn{1}{|>{\centering}p{1.4cm}}{} &  & \multicolumn{9}{>{\centering}p{10.8cm}|}{from}\tabularnewline
\cline{3-11} 
\multicolumn{1}{|>{\centering}p{1.4cm}}{} &  & L2/3e & L2/3i & L4e & L4i & L5e & L5i & L6e & L6i & Th\tabularnewline
\hline 
\multirow{8}{1.4cm}{\rotate{to}} & L2/3e & 0.101 & 0.169 & 0.044 & 0.082 & 0.032 & 0.0 & 0.008 & 0.0 & 0.0\tabularnewline
\cline{2-11} 
 & L2/3i & 0.135 & 0.137 & 0.032 & 0.052 & 0.075 & 0.0 & 0.004 & 0.0 & 0.0\tabularnewline
\cline{2-11} 
 & L4e & 0.008 & 0.006 & 0.050 & 0.135 & 0.007 & 0.0003 & 0.045 & 0.0 & 0.0983\tabularnewline
\cline{2-11} 
 & L4i & 0.069 & 0.003 & 0.079 & 0.160 & 0.003 & 0.0 & 0.106 & 0.0 & 0.0619\tabularnewline
\cline{2-11} 
 & L5e & 0.100 & 0.062 & 0.051 & 0.006 & 0.083 & 0.373 & 0.020 & 0.0 & 0.0\tabularnewline
\cline{2-11} 
 & L5i & 0.055 & 0.027 & 0.026 & 0.002 & 0.060 & 0.316 & 0.009 & 0.0 & 0.0\tabularnewline
\cline{2-11} 
 & L6e & 0.016 & 0.007 & 0.021 & 0.017 & 0.057 & 0.020 & 0.040 & 0.225 & 0.0512\tabularnewline
\cline{2-11} 
 & L6i & 0.036 & 0.001 & 0.003 & 0.001 & 0.028 & 0.008 & 0.066 & 0.144 & 0.0196\tabularnewline
\hline 
\multicolumn{2}{|@{\hspace*{1mm}}p{2.7cm}@{}|}{\textbf{Name}} & \multicolumn{3}{@{\hspace*{1mm}}p{3.3cm}@{}|}{\textbf{Value}} & \multicolumn{6}{@{\hspace*{1mm}}p{6.6cm}|}{\textbf{Description}}\tabularnewline
\hline 
\multicolumn{2}{|>{\raggedright}p{1.35cm}|}{$w\pm\delta w$} & \multicolumn{3}{@{\hspace*{1mm}}p{3.3cm}@{}|}{$87.8\pm8.8\pA$ } & \multicolumn{6}{@{\hspace*{1mm}}p{6.6cm}@{}|}{excitatory synaptic strengths}\tabularnewline
\hline 
\multicolumn{2}{|l|}{$g$} & \multicolumn{3}{@{\hspace*{1mm}}p{3.3cm}@{}|}{$-4$} & \multicolumn{6}{@{\hspace*{1mm}}p{6.6cm}@{}|}{relative inhibitory synaptic strength}\tabularnewline
\hline 
\multicolumn{2}{|l|}{$d_{\mathrm{e}}\pm\delta d_{\mathrm{e}}$} & \multicolumn{3}{@{\hspace*{1mm}}p{3.3cm}@{}|}{$1.5\pm0.75\ms$} & \multicolumn{6}{@{\hspace*{1mm}}p{6.6cm}@{}|}{excitatory synaptic transmission delays}\tabularnewline
\hline 
\multicolumn{2}{|l|}{$d_{\mathrm{i}}\pm\delta d_{\mathrm{i}}$} & \multicolumn{3}{@{\hspace*{1mm}}p{3.3cm}@{}|}{$0.8\pm0.4\ms$} & \multicolumn{6}{@{\hspace*{1mm}}p{6.6cm}@{}|}{inhibitory synaptic transmission delays}\tabularnewline
\hline
\end{tabular}

\begin{tabular}{|@{\hspace*{1mm}}p{3.1cm}@{}|@{\hspace*{1mm}}p{3.65cm}@{}|@{\hspace*{1mm}}p{8.1cm}|}
\hline 
\multicolumn{3}{|>{\columncolor{parametergray}}c|}{\textbf{D: Neuron Model}}\tabularnewline
\hline 
\textbf{Name} & \textbf{Value } & \textbf{Description}\tabularnewline
\hline 
$\taum$ & $10\ms$ & membrane time constant\tabularnewline
\hline 
$\tau_{\mathrm{ref}}$ & $2\ms$ & absolute refractory period\tabularnewline
\hline 
$\tau_{\mathrm{syn}}$ & $0.5\ms$  & postsynaptic current time constant\tabularnewline
\hline 
$C_{\mathrm{m}}$ & $250\pF$  & membrane capacity\tabularnewline
\hline 
$V_{\mathrm{reset}}$ & $-65\mV$ & reset potential\tabularnewline
\hline 
$\theta$ & $-50\mV$ & fixed firing threshold\tabularnewline
\hline 
$\nu_{\mathrm{th}}$ & $15\Hz$  & thalamic firing rate during input period\tabularnewline
\hline
\end{tabular}

\caption{Parameter specification. The categories refer to the model description
in \prettyref{tab:Model-description}. \label{tab:Simulation-parameters}}

\end{table}

\subsection*{Network simulations\phantomsection\addcontentsline{toc}{subsection}{Network simulations} }

The network is defined by eight neuronal populations representing
the excitatory and inhibitory cells in the four layers 2/3, 4, 5 and
6. The populations consist of current-based leaky integrate-and-fire
model neurons with exponential synaptic currents (see \emph{Supplemental
Material}) and are randomly connected with connection probabilities
according to the integrated connectivity map we derive in this article.
Every population receives Poissonian background spike trains \citep{Amit97,Brunel00_183};
the firing rates of these inputs are composed by the number of external
inputs a neuron in a particular population receives and the background
spike rate contributed by each synapse. Synaptic strengths and synaptic
time constants of all connections are chosen such that an average
excitatory postsynaptic potential (EPSP) has an amplitude of $0.15\mV$
with a rise time of $1.6\ms$ and a width of $8.8\ms$ mimicking the
\emph{in vivo} situation \citep{Fetz91}. Inhibitory strengths are
negative and increased by a factor $g$ compared to the excitatory
ones. Delays in the network are chosen independent of the layer, with
excitatory delays being on average around twice as long as inhibitory
delays but the exact ratio is uncritical. To introduce heterogeneity
into the network, we draw the synaptic strengths and delays from Gaussian
distributions (prohibiting a change of sign of the strengths and constricting
delays to be positive and multiples of the computation step size).
The network structure and a complete list of parameters and their
values for the reference network model are systematically described
according to \citet{Nordlie-2009_e1000456} in Tables \ref{tab:Model-description}
and \ref{tab:Simulation-parameters}.

In some cases, we explicitly model the thalamic input to L4 and L6
by a thalamic population of 902 neurons \citep{Peters93_69} that
emit Poissonian spike trains at a given rate in some prescribed time
interval. These relay cells are randomly connected to the cortex with
the cell-type specific connection probabilities according to \citet{Binzegger04},
see \prettyref{tab:Simulation-parameters}.

To instantiate the network model we randomly draw for every synapse
the pre- and the postsynaptic neuron. In contrast to the often used
convergent and divergent connectivity schemes \citep{Eppler09_12},
this procedure results in binomially distributed numbers of incoming
and outgoing synapses. In practice, we could first calculate the total
number of synapses forming a connection by inverting \prettyref{eq:connection_prob}
and then successively create the synapses. In a distributed simulation
setup, however, this procedure is inefficient because the neurons
are distributed over multiple processes. Although a synapse is only
created if the postsynaptic neuron is local to the process \citep{Morrison05a}
the full algorithm would have to be carried out on each process. We
solve this problem by calculating a priori how many synapses will
be created locally on each process, exploiting that the distribution
of synapses over processes is multinomial. Subsequently we apply the
serial algorithm on every machine to the local synapses only: The
presynaptic neuron is drawn from all neurons in the presynaptic population
and the postsynaptic cell on a given process is drawn only from the
neurons located on this process (compare supplementary Fig. 16). While
the first step is serial, but efficient for the number of processes
we typically employ, the second step is fully parallelized. The procedure
is detailed in the \emph{Supplemental Material}. A reference implementation
of this algorithm will be made available in the NEST simulation tool
(\url{http://www.nest-initiative.org}) as RandomPopulationConnect.

All simulations are carried out with the NEST simulation tool \citep{Gewaltig_07_11204}
using a grid constrained solver and a computation step size $h=0.1\ms$
on a compute cluster with 24 nodes each equipped with 2 quad core
AMD Opteron 2834 processors and interconnected by a 24 port Voltaire
InfiniBand switch ISR9024D-M. 48 cores simulate the network of around
80,000 neurons and 0.3 billion synapses in close to real-time \citep{Djurfeldt10_43}.

\section*{Results\phantomsection\addcontentsline{toc}{section}{Results} \label{sec:Results}}

\begin{figure}
\begin{centering}
\includegraphics{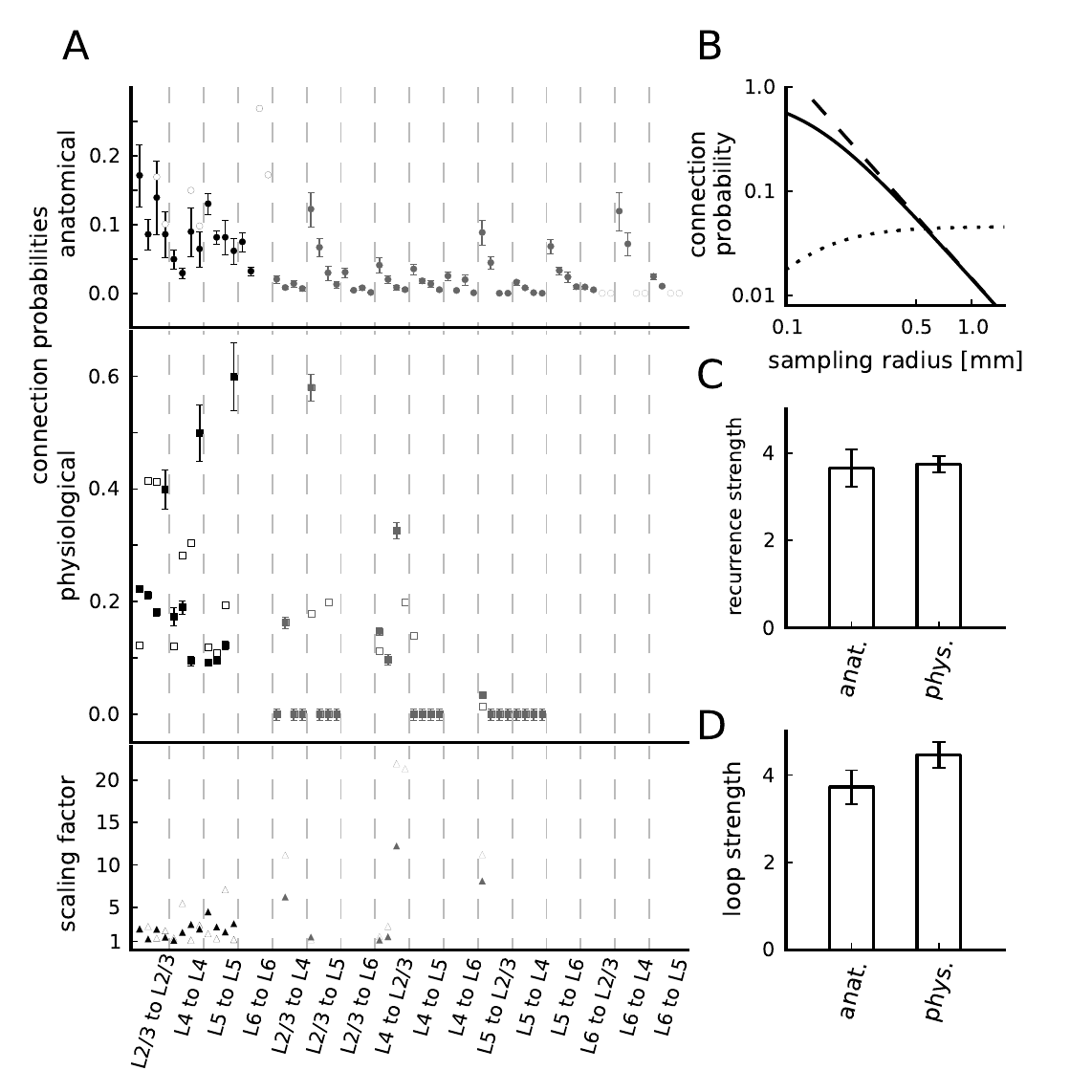}
\par\end{centering}

\caption{Properties of connectivity maps. (A) Connection probabilities according
to the anatomical (top panel, circles) and physiological connectivity
map (center panel, squares) and corresponding scaling factors (bottom
panel, triangles). Both, raw (closed markers) and modified (open markers)
are shown. The data are horizontally arranged according to their classification
as within-layer (black) and inter-layer (gray) connections. For a
given pair of pre- and post-synaptic layer, the data are arranged
from left to right according to connection types: excitatory to excitatory,
excitatory to inhibitory, inhibitory to excitatory and inhibitory
to inhibitory. L5i to L5e/i outside of the displayed range, see supplementary
Table 6. Error bars are minimal statistical errors (see \emph{Supplemental
Material}). (B) Sampling radius dependence of anatomical connection
probability (solid: \prettyref{eq:connection_prob}, dashed: \prettyref{eq:approx_conn_prob})
and of the product of connection probability and area (dotted) in
double-logarithmic representation. (C) Anatomical and physiological
recurrence strength. (D) Anatomical and physiological loop strength.
Error bars in C and D are based on the minimal statistical error estimates
of connection probabilities using error propagation.}

\label{fig:connmatrixproperties}
\end{figure}

\subsection*{Comparison of connectivity maps\phantomsection\addcontentsline{toc}{subsection}{Comparison of connectivity data}\label{sub:Res-Comparison-of-connectivity}}

The anatomical and the physiological connectivity maps exhibit a similar
structure (\prettyref{fig:connmatrixproperties}): Recurrent within-layer
connections are all non-zero with connection probabilities tending
to decrease from superficial to deeper layers, excepting the physiological
interneuron to interneuron connections which are, however, subject
to poor statistics (\prettyref{tab:Physiological-data}). The inter-layer
connections can be subdivided into connections with similar probabilities
as within-layer connections and those with values close or equal to
zero. Consequently, the recurrence strength, quantifying the relative
strength of recurrent within-layer connections, is greater than one
and statistically indistinguishable for the two maps (z-test, $P>0.1$,
\prettyref{fig:connmatrixproperties}C). Similarly, the loop strength,
comparing the connectivity of the feed-forward loop to all other inter-layer
connections, is indistinguishable (z-test, $P>0.05$, \prettyref{fig:connmatrixproperties}D).
However, 50\% of the scaling factors of inter-layer connections are
large (L2/3e to L4i, L5e to L2/3e, L4i to L2/3e and L4i to L2/3i)
for both, the raw and the modified data.

\subsection*{Lateral connectivity \phantomsection\addcontentsline{toc}{subsection}{Lateral connectivity}\label{sub:Res-model-of-lateral}}

In contrast to the largely consistent relative measures (\prettyref{fig:connmatrixproperties}C
and D), absolute connection probability values differ (supplementary
Table 6). We hypothesize that the differences in the mean connection
probabilities are explained by differences in the methodology applied
to obtain the connectivity maps: Physiological recordings in slices
are usually restricted to a maximal lateral distance of the somata
of around $100\mum$ (as reported in \citet{Thomson02} for the raw
physiological map). The anatomical data, in contrast, are based on
reconstructed axons and dendrites, with axons extending up to $4\mm$
\citep{Binzegger07_12242}, in general beyond $1\mm$. When providing
absolute numbers, \citet{Binzegger04} refer to the surface area of
cat area 17 ($399\mm^{2}$).

\begin{figure}
\begin{centering}
\includegraphics[width=8.5cm]{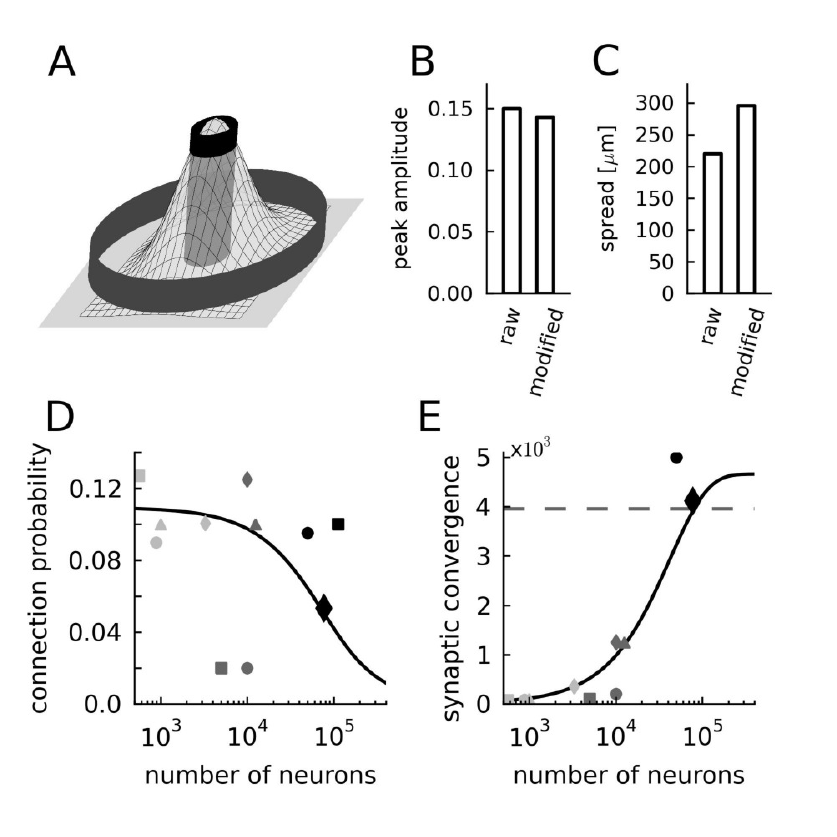}
\par\end{centering}

\caption{Lateral connectivity model. (A) Two dimensional Gaussian with two
cylinders indicating the lateral sampling of the anatomical (gray
outer cylinder) and the physiological (black inner cylinder) experiments.
(B) Estimated peak amplitude and (C) lateral spread of the connectivity
model based on mean connectivity of the anatomical and physiological
raw and modified maps. (D) Average connection probability and (E)
average synaptic convergence (average number of synaptic inputs per
neuron) of the layered network model as a function of the network
size expressed in number of neurons. The dashed horizontal line marks
85\% of the maximal synaptic convergence in the local network. Black
diamonds show the data used in our simulations, further markers indicate
other published cortical network models: \citet{Haeusler07_149} (light
gray square), \citet{Izhikevich06_245} (light gray triangle), \citet{Izhikevich04}
(light gray circle, embedded local network is defined by the area
receiving connections from a single long-range axon), \citet{Lundqvist06_253,Djurfeldt08_31}
(light gray diamond, local network represented by one hypercolumn),
\citet{Vogels05b,Vogels05a} (dark gray circle), \citet{Sussillo07_4079}
(dark gray square), \citet{Brunel00_183} (dark gray triangle), \citet{Kriener08_2185}
(dark gray diamond), \citet{Kumar08_1} (black circles), \citet{Morrison07_1437}
(black square).\label{fig:Lateral-connectivity-model}}

\end{figure}

We employ a Gaussian lateral connectivity model (\prettyref{eq:Gauss}),
similar to the one by \citet{Hellwig00_111,Buzas06} to account for
the different experimental sampling radii (see \emph{Materials and
Methods}). We assume the model to reflect the \emph{in vivo} connectivity
structure and the experimental connectivity maps to characterize samples
of this structure. The anatomical measurement is interpreted as an
unconstrained sampling over the complete lateral connectivity structure
while the physiological measurement corresponds to a local measure
sampling from the center region of the Gaussian (\prettyref{fig:Lateral-connectivity-model}A). 

The model parameters, peak connection probability and lateral spread,
are determined based on the mean connection probabilities of the two
maps and the physiological sampling radius (eqs. \eqref{eq:sigma}
and \eqref{eq:c0}, \prettyref{fig:Lateral-connectivity-model}).
The estimated lateral spread is consistent with data from rat and
cat primary visual cortex based on the potential connectivity method:
\citet{Hellwig00_111}, caption to his figure 7, reports a lateral
spread of 150 to 310 $\mu$m, \citet{Stepanyants07}, their figure
7, find a spread of around 200 $\mu$m of main projections in input
and output maps. Also the overall connectivity level of $0.138$ for
nearby neurons with a distance of $100\mum$ is in good agreement
with the extensively used estimate of $0.1$ provided by \citet{Braitenberg98}.
These consistencies indicate that our underlying assumption --anatomical
and physiological experiments sample independently from the same lateral
connectivity profile-- is valid.

\subsubsection*{Average model connectivity}

Our network model does not incorporate a lateral connectivity structure
but randomly connected populations. We use the Gaussian lateral connectivity
model exclusively to determine the average connection probability
for a given network size (\prettyref{eq:cm3}, \prettyref{fig:Lateral-connectivity-model}D).
The connectivity of small networks (up to about 7,000 neurons) is
largely determined by the physiological connectivity and the one of
large networks (above 100,000 neurons) by the anatomical connection
probability (decaying quadratically, see \prettyref{fig:connmatrixproperties}B).
For intermediate network sizes, \prettyref{eq:cm3} interpolates between
these two extremes according to the Gaussian lateral connectivity
profile. \prettyref{fig:Lateral-connectivity-model}E shows the average
synaptic convergence as a function of the network size. It reveals
that only for large networks the majority of local synapses is represented
in the model: for example a network of around 80,000 neurons comprises
more than 85\% of all local synapses. In contrast, a network of e.g.
10,000 neurons represents only around 20\% of the local connectivity.
Therefore, we select our network to correspond to $1\mm^{2}$ of cortical
surface (77,169 neurons).

According to our analysis the maximal average number of synapses per
neuron is about 5,000. This number is consistent with the data of
\citeauthor{Binzegger04}\citeyearpar[ see their figure 11A]{Binzegger04}.
In \prettyref{fig:Lateral-connectivity-model} we also display the
connectivities and convergences of a selection of other modeling studies
on the local cortical network (some data representing local networks
embedded in larger networks). Independent of model size most studies
use a connection probability around $0.1$ which is largely consistent
with our results. Only for large networks of 50,000 neurons and more,
this connection probability is above our estimate. Two studies use
a significantly smaller connection probabilities of $0.02$, arguing
that this value interpolates between high local and low distal connectivity.
Although the absolute numbers differ from our estimates, the reasoning
is the same as for our model. In all but two cases \citep{Morrison07_1437,Kumar08_1}
the models' convergence is maximally 20\% of the anatomical estimate.

\subsection*{Randomness and specificity\phantomsection\addcontentsline{toc}{subsection}{Randomness and specificity}\label{sub:Res-Target-specificity}}

\begin{figure}
\begin{centering}
\includegraphics{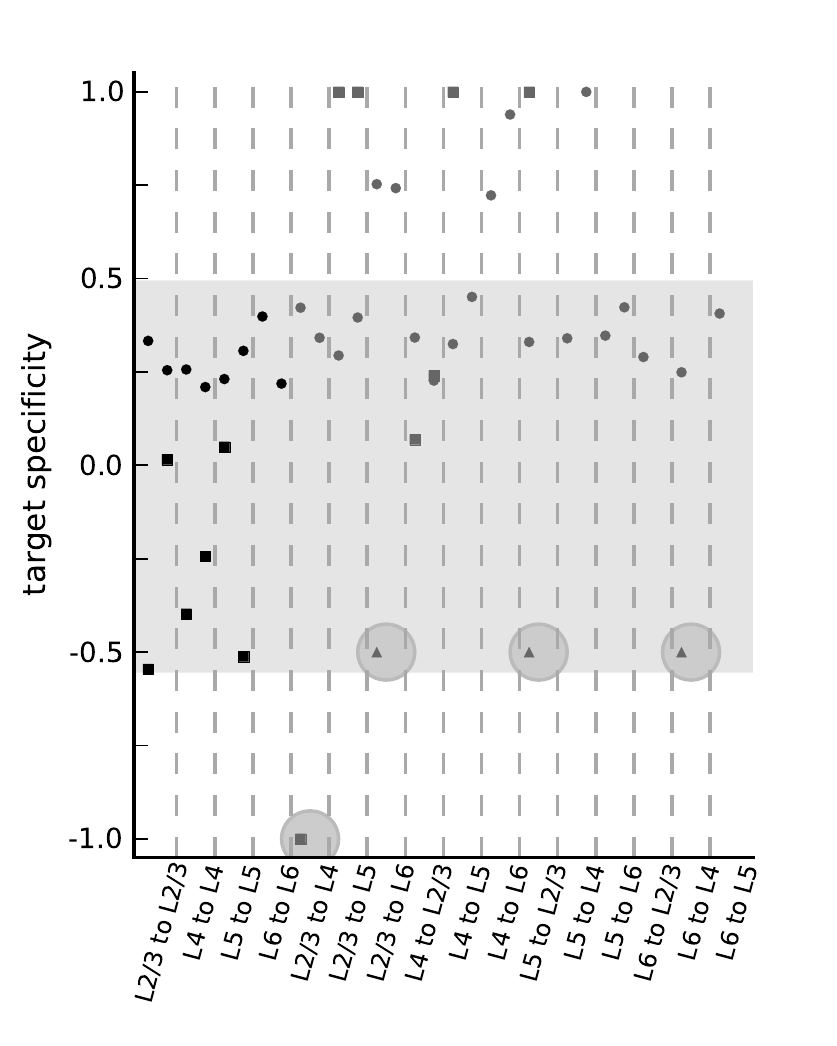}
\par\end{centering}

\caption{Anatomical (circles) and physiological (squares) target specificity
estimates based on the modified connectivity maps. $+1(-1)$ indicates
exclusive selection of excitatory (inhibitory) targets, $0$ random
selection. Triangles show additional candidates of inhibition-specific
projections discovered in photostimulation or electron microscopy
studies. The shaded area highlights the estimates of within-layer
projections and inter-layer projections with target specificity values
between $0$ and $0.5$ ({}``non-specific'' connections). Within
these bounds the anatomical and physiological estimates are $0.32\pm0.07$
and $-0.17\pm0.28$, respectively. The largest statistically well
constrained values from a single laboratory are exhibited by the projections
L2/3e to L2/3, L4e to L4 and L5e to L5 (compare supplementary Table
6) of the raw physiological map. The target specificity of these data
is $-0.01\pm0.03$. The four highlighted data points are candidates
of specific target type selection. For every pair pre- and post-synaptic
layer the figure shows data both pre-synaptic neuron types (left:
excitatory, right: inhibitory).\label{fig:Target-specificity}}

\end{figure}

A central assumption of the anatomical connectivity map is randomness,
i.e. synapses are established independent of the pre- and postsynaptic
cell type. Nevertheless, the target specificity estimates of the anatomical
connectivity map (circles in \prettyref{fig:Target-specificity})
are $>0.2$, reflecting a preferential selection of excitatory targets.
The bias is introduced by the application of Peters' rule, which assumes
synaptic densities on dendrites to be independent of cell-type, to
bouton densities and dendritic lengths. Furthermore, some projections
exhibit very high values $>0.5$ because only the dendrites of excitatory
cells, but barely those of inhibitory cells, reach into the cloud
of presynaptic axonal elements.

The target specificity values of the physiological map contrast with
the anatomical findings. Most values (squares in \prettyref{fig:Target-specificity})
are consistently smaller and show larger variability than the anatomical
estimates. In addition, excitatory within-layer connections of the
raw connectivity map univocally select their targets independent of
the postsynaptic type. Several projections in the physiological map,
however, connect exclusively to excitatory neurons due to incomplete
sampling of inhibitory subtypes. This highlights that a straightforward
application of the physiological connectivity map in simulations results
in artifacts due to missing feed-forward inhibition. Finally, the
projection from L2/3e to L4 specifically targets inhibitory, but not
excitatory cells (see also \prettyref{tab:Physiological-data}).

The specific selection of inhibitory targets of the L2/3e to L4 projection
cannot be explained by differences in the overlap of the excitatory
and inhibitory dendrites with the excitatory axons and is therefore
beyond the scope of anatomical studies relying on the statistics of
neuronal morphology such as Peters' rule \citep{Binzegger04} and
also potential connectivity \citep{Stepanyants07}. The specificity
in the inter-layer circuitry explains the large scaling factor of
this projection (\prettyref{fig:connmatrixproperties}A).

We identify three additional candidates of specific target type selection:
L5e to L2/3 (\citet{Dantzker00,Thomson98_669,Thomson02_936,Lefort09_301})
L2/3e to L6 \citep{Zarrinpar06_1751} and L6e to L4 \citep{McGuire84_3021},
see\emph{ Supplemental Material} for details on the choice of these
candidates. The structure of these projections has not been quantified
comprehensively in paired recordings and evidence suggests that the
target type selection is less strict than for the L2/3e to L4 projection.
We tentatively assume them to have intermediate values (triangles
in \prettyref{fig:Target-specificity}). Two of these projections
(L2/3e to L4 and L5e to L2/3) are inverse to the feed-forward loop.
\citet{Thomson02,Thomson02_1781} argued that the specific target
type selection plays a distinct functional role because the inhibition-specific
({}``i-specific'') feedback projections may prevent reverberant
excitation involving L2/3, L4 and L5 and enhance the propagation of
synchronous thalamic inputs.

We utilize the information on i-specific feedback and the anatomical
estimates to remove for some projections the methodological biases
by consistently modifying the respective target specificities (see
\emph{Materials and Methods} and \prettyref{tab:Amendment-candidates-for-tsp}).
Thereby, we estimate previously not measured physiological connection
probabilities and introduce the specific selection of targets into
the anatomical map. The latter constitutes an effective redistribution
of synapses and corresponds to a refinement of Peter's rule (supplementary
Fig. 13).

\begin{table}
\begin{centering}
\begin{tabular}{lcl}
projection & $T$ & data source\tabularnewline
\hline
\hline 
L2/3e to L4 & -0.8 & \citealt{Thomson02_936}\tabularnewline
L5e to L2/3 & -0.4 & \citealt{Dantzker00}\tabularnewline
L2/3e to L6 & -0.4 & \citealt{Zarrinpar06_1751}\tabularnewline
L6e to L4 & -0.4 & \citealt{McGuire84_3021}\tabularnewline
L2/3e to L5 & 0.29 & \citealt{Binzegger04}\tabularnewline
L4e to L5 & 0.32 & \citealt{Binzegger04}\tabularnewline
L2/3i to L5 & 0.4 & \citealt{Binzegger04}\tabularnewline
L4i to L2/3 & 0.23 & \citealt{Binzegger04}\tabularnewline
\end{tabular}
\par\end{centering}

\caption{Amendment candidates for target specificity. The rows describe the
projections whose target specificities are modified during the compilation
of the integrated connectivity map. For each projection the second
column states the target specificity value $T$ after the amendment,
the third the publication on which the modification is based. The
top four rows are the candidate projections for a preference of inhibitory
targets. In these cases, no quantitative estimates are known. $T=-0.8$
is set if the literature provides a strong indication and $T=-0.4$
for a comparably weak indication (compare \emph{Supplemental Material}).
The $T$-values of the bottom four rows are based on the anatomical
map and provide estimates of the previously not measured connections
to inhibitory neurons for the physiological map.\label{tab:Amendment-candidates-for-tsp}}

\end{table}

\subsection*{Integrated connectivity map\phantomsection\addcontentsline{toc}{subsection}{Integrated connectivity map}\label{sub:Res-Integrated-connectivity-data}}

Based on the information gathered in the previous sections we now
compile an integrated connectivity map. Our algorithmic compilation
(see \emph{Materials and Methods}) accounts for the differences in
the lateral sampling of the two methods and corrects for methodological
shortcomings expressed in the target specificity of projections (\prettyref{tab:Amendment-candidates-for-tsp})
by incorporating photostimulation and electron microscopy data.

\begin{figure}
\begin{centering}
\includegraphics{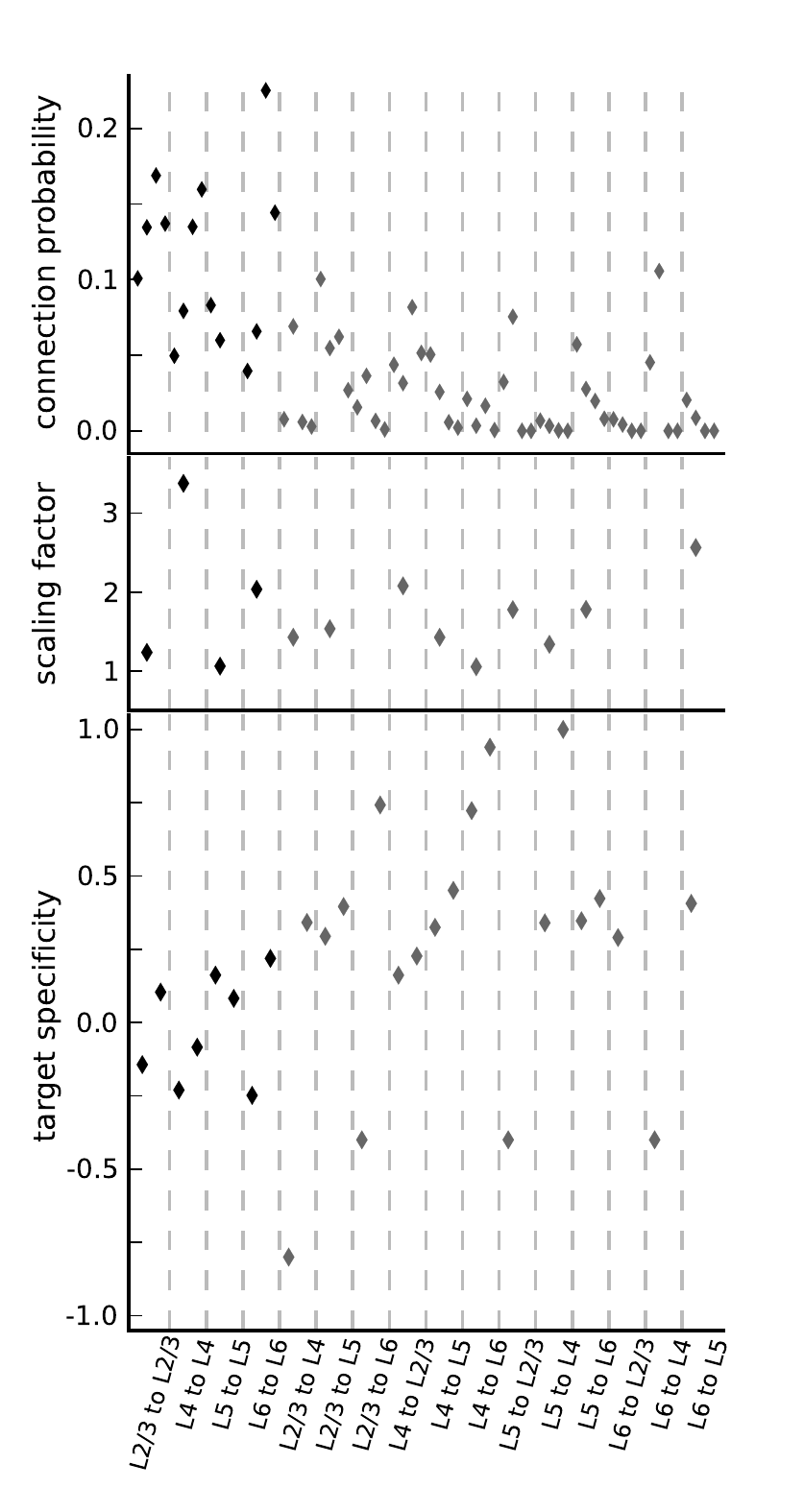}
\par\end{centering}

\caption{Properties of the integrated connectivity map. Layout of graphs as
in \prettyref{fig:connmatrixproperties} and \prettyref{fig:Target-specificity}
respectively. The top panel shows the connection probabilities of
the model (L5i to L5e/i outside the displayed range, see \prettyref{tab:Simulation-parameters}).
The center panel shows the scaling factors between the integrated
map, restricted to excitatory to excitatory connections, and the recently
published map of \citet{Lefort09_301}. The bottom panel shows the
target specificity of the integrated connectivity map (within-layer
connections: $-0.02\pm0.17$, non-specific inter-layer connections,
see \prettyref{fig:Target-specificity}: $0.33\pm0.08$).\label{fig:integrated-data-conn-tsp}}

\end{figure}

The resulting connection probabilities are shown in the upper panel
of \prettyref{fig:integrated-data-conn-tsp}. Visual inspection shows
that the overall structure of the original connectivity maps is preserved.
For a consistency check, we calculate the scaling factors between
our integrated map and the recently published data for excitatory
to excitatory connections of mouse C2 barrel column \citep{Lefort09_301}.
In this comparison we combine the data of \citeauthor{Lefort09_301}
on L2 and L3 to L2/3 and on L5A and L5B to L5 according to \prettyref{eq:conn_prob_paired}.
The scaling factors (see \prettyref{fig:integrated-data-conn-tsp},
center panel) are low, indicating a good agreement of the connectivity
maps. The main outlier is the recurrent L4e to L4e connection, which
exhibits the highest connection probability in the study of \citet{Lefort09_301}
but rather low values in the physiological and especially the anatomical
map. The lower panel of \prettyref{fig:integrated-data-conn-tsp}
shows the target specificity values of the integrated connectivity
map. Within-layer projections select their targets randomly and inter-layer
projections inherit mostly the properties of the anatomical map, except
for the four candidate i-specific projections.

\begin{figure}
\begin{centering}
\includegraphics{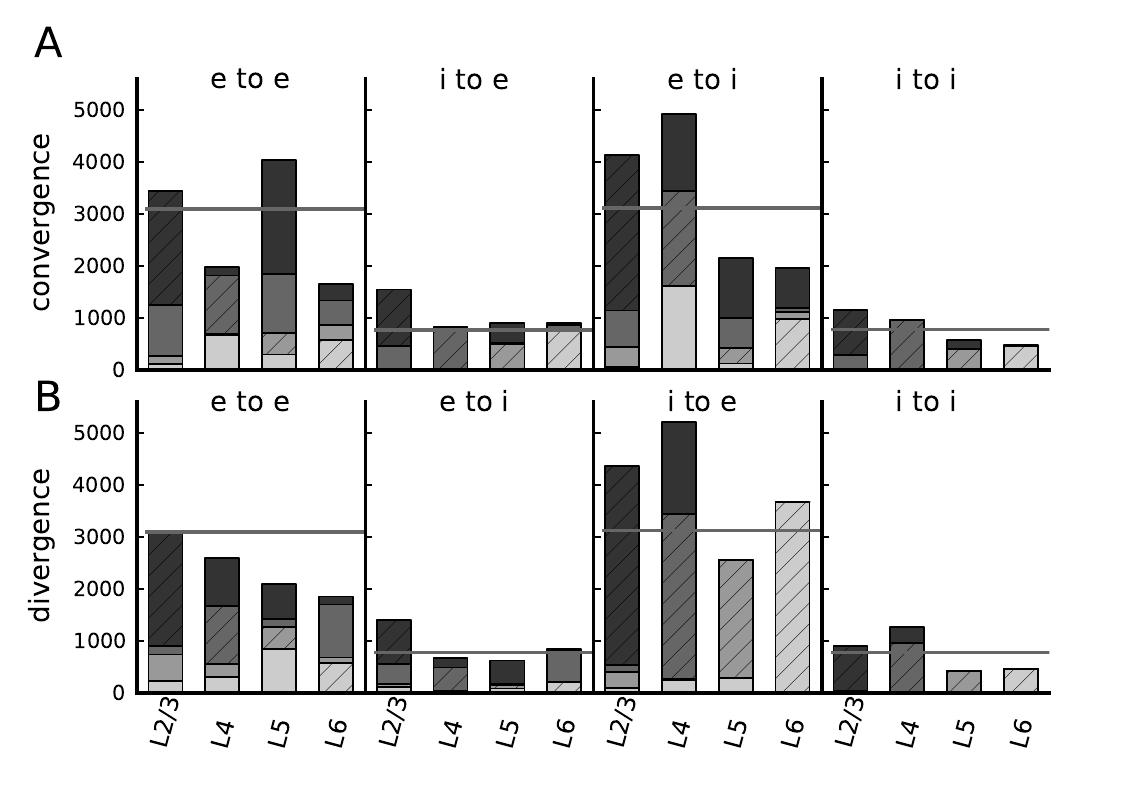}
\par\end{centering}

\caption{Cell-type specific convergence (A) and divergence (B) of the integrated
connectivity map. The histograms display blocks of data for the four
different connection types between excitatory (e) and inhibitory (i)
neurons (as indicated). For a neuron in the layer specified on the
horizontal axis, the individual bar segments show the absolute number
of synapses the neuron receives from a source layer (convergence)
or establishes in a target layer (divergence). Bar segments are arranged
according to the physical location of the layers in the cortex (from
top to bottom: L2/3, L4, L5, L6). Hatched bars represent within-layer
connections. Lightness of gray increases from superficial to deeper
layers. Gray horizontal lines indicate the convergence and divergence
of a balanced random network with the same total number of neurons
and synapses as the layered model.\label{fig:integrated-convergence-divergence}}

\end{figure}

The cell-type specific convergences and divergences (\prettyref{fig:integrated-convergence-divergence})
show that the integrated connectivity map reflects prominent features
of local cortical connectivity: Except for L5e, convergence is dominated
by within-layer connections \citep[consistent with e.g.][]{Douglas91a,Douglas04}.
Furthermore, the strongest inter-layer excitatory to excitatory divergences
correspond to the feed-forward loop from L4 to L2/3 to L5 to L6 and
back to L4 \citep{Gilbert83_217,Gilbert83}. The excitatory to inhibitory
divergence is dominated by the i-specific feedback connections.

By comparing the average convergence and divergence to the neuronal
densities of the different cell populations (supplementary Fig. 14)
we find that neurons in the local microcircuit sample most excitatory
inputs from L2/3 and fewest from L5 and L6. In contrast, local outputs
within the microcircuit project preferentially to L5.

\subsubsection*{External inputs}

The model consists in total of about 217 million excitatory and 82
million inhibitory synapses. The inhibitory synapse count is consistent
with \citet{Beaulieu85} ($64\pm21$ million) while the number of
excitatory synapses is lower than their estimate ($339\pm43$ million)
presumably reflecting that a fraction of all excitatory synapses originates
outside of the local network. The ratio of local synapses (total number
of synapses in our network model) and all synapses (according to the
countings of \citet{Beaulieu85}) is 0.74, similar to the ratio reported
by \citet{Binzegger04}, but see \citet{Stepanyants09_3555}.

The layer-specific external input structure distinguishes between
thalamic afferents, gray-matter and white-matter inputs (see \emph{Materials
and Methods}). The resulting total number of external inputs per neuron
(\prettyref{tab:Simulation-parameters}) is lowest in L2/3, intermediate
in L4 and L5 and highest in L6. This relative structure is largely
dominated by the white-matter inputs. In contrast, most gray-matter
long-range inputs form synapses on L2/3 neurons and fewest form synapses
in L6. Given the presently available data, we cannot exclude that
some synapses treated here as external input actually represent a
pathway in the local microcircuit still awaiting comprehensive experimental
assessment.

\subsection*{Spontaneous layer-specific activity\phantomsection\addcontentsline{toc}{subsection}{Spontaneous layer-specific activity}}

\begin{figure}
\begin{centering}
\includegraphics{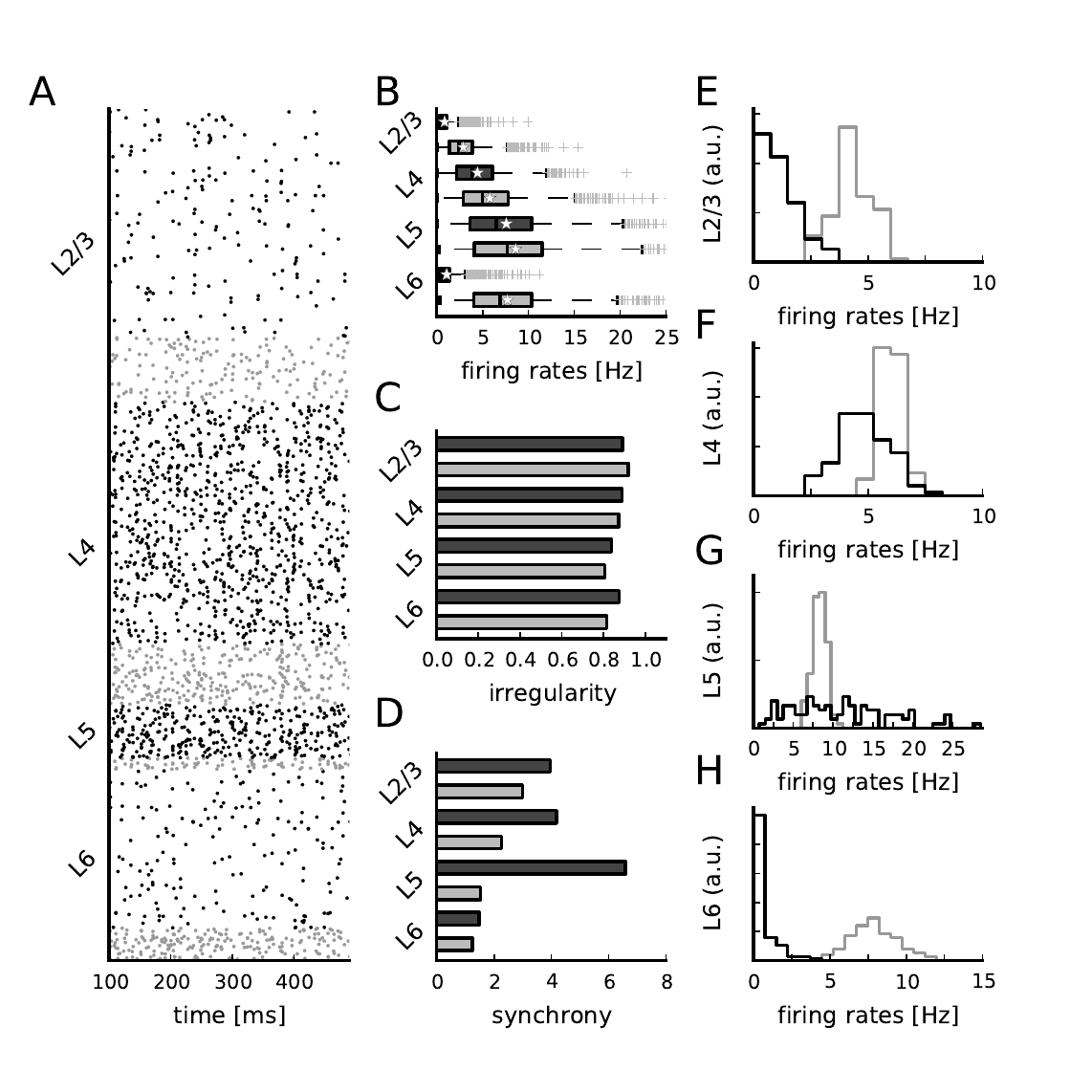}
\par\end{centering}

\caption{Spontaneous cell-type specific activity. (A) Raster plot of spiking
activity recorded for 400 ms of biological time of layers 2/3, 4,
5 and 6 (from top to bottom; black: excitatory, gray: inhibitory).
Relative number of displayed spike trains corresponds to the relative
number of neurons in the network (total of 1862 shown). (B-D) Statistics
of the spiking activity of all eight populations in the network based
on 1000 spike trains recorded for $60\s$ (B and C) and $5\s$ (D)
for every population. (B) Boxplot \citep{Tukey77} of single-unit
firing rates. Crosses show outliers, stars indicate the mean firing
rate of the population. (C) Irregularity of single-unit spike trains
quantified by the coefficient of variation of the inter-spike intervals.
(D) Synchrony of multi-unit spiking activity quantified by the variance
of the spike count histogram (bin width $3\ms$) divided by its mean.
(E-H) Histograms of the excitatory and inhibitory population firing
rates for L2/3 (E), L4 (F), L5 (G) and L6 (H) for randomly drawn external
inputs (100 trials, see \emph{Materials and Methods} for details).
\label{fig:stationary-activity}}

\end{figure}

The simulated spontaneous spiking activity of all cell types corresponds
to the asynchronous irregular activity state observed in mono-layered
balanced random network models \citep{Amit97,Brunel00_183}. \prettyref{fig:stationary-activity}
(A-D) shows the ongoing spontaneous spiking activity of all populations
and the corresponding firing rates, irregularity and synchrony. The
activity varies significantly across cell types. L2/3e and L6e exhibit
the lowest firing rates with a mean below or close to $1\Hz$. L4e
cells fire more rapidly at around $4\Hz$ and L5e cells at more than
$7\Hz$. In all layers, inhibitory firing exceeds excitatory rates.
The boxplots furthermore visualize that the firing rates of single
neurons can differ substantially. For instance in L2/3e, several neurons
fire at more than $5$ Hz while the majority of neurons is rather
quiescent emitting less than one spike per second. This effect is
due to the binomially distributed convergence.

Single unit activity is irregular; the mean of the single unit coefficients
of variation of the inter-spike intervals is greater than $0.8$.
The population activity is largely asynchronous, but exhibits fast
oscillations with low amplitude similar to balanced random networks
\citep[e.g. ][]{Brunel00_183}. We assess the synchrony of every population's
multi-unit activity by the variability of the spike count histogram
(\prettyref{fig:stationary-activity}D). At the given firing rates
and bin width, the synchrony of the spiking activity is highest in
L5e and lowest in L6. The synchrony of the membrane potential traces
(\citealp{Golomb07_1347}, supplementary Fig. 20) confirms that the
activity is asynchronous.

The observed activity features of the network do not depend on the
specific structure of the external inputs: replacing the Poissonian
background by a constant current (DC) to all neurons (supplementary
Fig. 17) or applying layer-independent Poissonian background inputs
(supplementary Fig. 18) yields similar results. Random selection of
the number of external inputs (see \emph{Materials and Methods}) also
confirms the previous findings: \prettyref{fig:stationary-activity}
(E-H) shows the histograms of the population firing rates in the different
layers for excitatory and inhibitory cells, respectively. The distribution
of excitatory population firing rates in L2/3 and L6 are persistently
low. L5e activity exhibits the highest firing rates and also the largest
variability. L4e and the inhibitory populations vary only slightly
with mean firing rates similar to the reference parametrization (\prettyref{fig:stationary-activity}B).
Apparently, the local microcircuitry reconfigures the firing rate
distributions for different input situations while conserving general
features like the low rate regime in L2/3e and L6e. In 85\% of the
simulations, L2/3e and L6e fire at a lower rate than L4e and simultaneously
L5e exhibits the highest firing rate. Also, we observe the inhibitory
firing rate within a given layer to be higher than the excitatory
rate in 83\% of all cases.

\subsubsection*{Comparison to \emph{in vivo} activity}

\prettyref{tab:Comparison-firing-rates} contrasts the experimentally
observed firing rates in individual layers with our simulation results.
Experimentally, the spontaneous activity of L2/3e pyramids has been
studied extensively. Consistent over species and areas the firing
rate is below $1\Hz$, in good quantitative agreement with the model.
The L6e firing rates are similar to the model values, although the
experimental data base is more sparse. The L4e and L5e firing rates
of rat primary somato-sensory cortex are lower than in the model,
but L5e consistently shows the highest spontaneous activity, also
in rat auditory cortex. The activity of cortico-tectal cells in L5
(and L4) of various cortices in the rabbit is slightly higher. Several
studies provide data on putative interneurons \citep{Swadlow88_1162,Swadlow89_288,Swadlow91_1392,Swadlow94_437,Sakata09_404},
demonstrating that inhibitory activity is typically higher than that
of excitatory cells. The statistics of single neuron spike trains
in our model show great variations due to the random connectivity
which also imposes a variance in the convergence of inputs. Therefore,
{}``neighboring'' neurons, i.e. neurons with statistically identical
connectivity, can exhibit very different firing rates, which has also
been observed experimentally \citep[e.g.][]{Swadlow88_1162,Heimel05_3538,deKock09_16446}.

\begin{sidewaystable}
\begin{centering}
\begin{tabular}{lllllll}
 &  & \multicolumn{4}{c}{Firing rates {[}$\mathrm{Hz}${]}} & \tabularnewline
\cline{3-6} 
Species & Area & L2/3e & L4e & L5e & L6e & data source\tabularnewline
\hline
\hline 
Mouse & S1 & $0.61$ & -- & -- & -- & \citet{Crochet06_608,Poulet08_881}\tabularnewline
Rat & V1 & $0.44$ & -- & -- & -- & \citet{Greenberg08_749}\tabularnewline
Rat & M1 & $0.36$ & -- & -- & -- & \citet{Lee06_399}\tabularnewline
Rat & S1 & $0.3$ & $1.4$ & $2$-$3$ & $0.5$ & \citet{deKock09_16446}\tabularnewline
Rat & A1 & $<1$ & -- & $3$-$4$ & -- & \citet{Sakata09_404}\tabularnewline
Rabbit & 4 areas & $<1$ & -- & $4$-$7$ & $<1$ & \citet{Swadlow88_1162,Swadlow89_288,Swadlow91_1392,Swadlow94_437}\tabularnewline
Squirrel & V1 & $0.19$ & $0.35$ & $1.7$ & $0.74$ & \citet{Heimel05_3538}\tabularnewline
Model & reference & $0.86$ & $4.45$ & $7.59$ & $1.09$ & \prettyref{fig:stationary-activity}B\tabularnewline
Model & random & $1.11\pm0.8$ & $4.8\pm1.1$ & $11\pm6.1$ & $0.56\pm0.9$ & \prettyref{fig:stationary-activity}E-H\tabularnewline
\end{tabular}
\par\end{centering}

\caption{Experimentally measured and simulated cell-type specific firing rates.
Numerical columns show the layer-resolved mean firing rates obtained
in \emph{in vivo }awake animal recordings and in the present modeling
study (last row indicates additionally the standard deviation). The
investigated areas are S1: primary somato-sensory cortex, V1: primary
visual cortex, M1: primary motor cortex, A1: primary auditory cortex.
The four areas investigated in the rabbit by \citet{Swadlow88_1162,Swadlow89_288,Swadlow91_1392,Swadlow94_437}
are V1, S1, S2 (secondary somato-sensory cortex) and M1, respectively.
In these four studies, L5 corresponds to cortico-tectal cells that
are partly also located in L4. The data for the gray squirrel are
the medians. The model results refer to the reference parametrization
and to the mean$\pm$std of the population rates for varying external
inputs. \label{tab:Comparison-firing-rates}}

\end{sidewaystable}

\subsubsection*{Stability of the network activity}

\begin{figure}
\centering{}\includegraphics{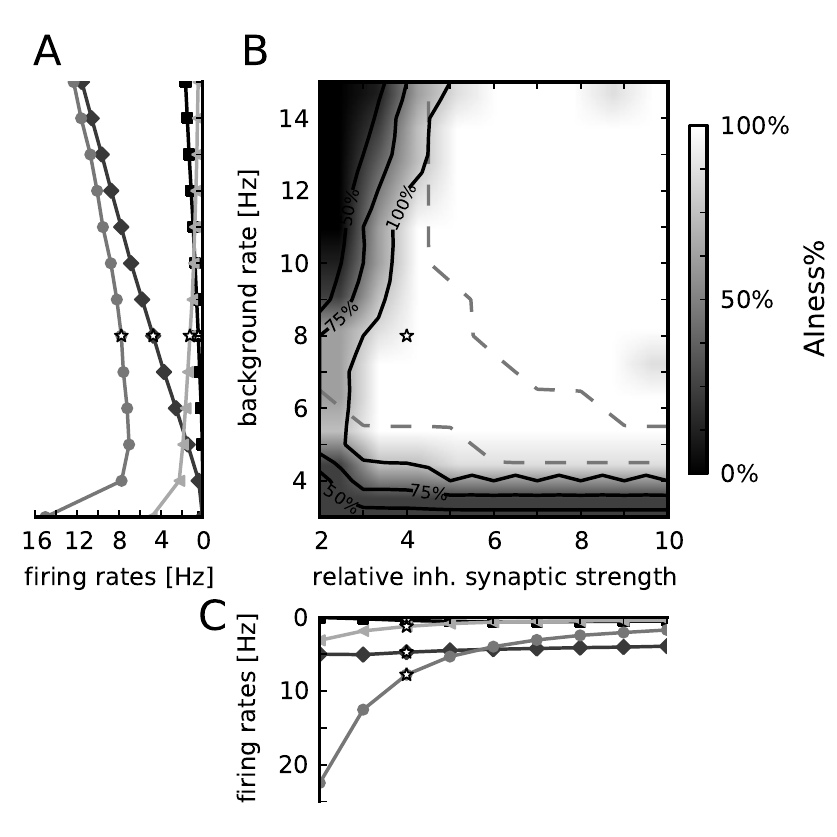}\caption{Dependence of network activity on the external background firing rate
and the relative inhibitory synaptic strength. White stars mark the
reference parameter set. (A) Population firing rates of excitatory
populations in layers 2/3 (squares), 4 (diamonds), 5 (circles) and
6 (triangles), lightness increases with cortical depth, as a function
of the background rate with fixed relative inhibitory synaptic strength
($4$). (B) AIness\%, the percentage of populations with firing rate
below $30$ Hz, irregularity between $0.7$ and $1.2$, and synchrony
below $8$ (data collected for $5\s$ per simulation), as a function
of the background rate and the relative inhibitory synaptic strength.
Labeled black iso-lines indicate areas where 50\%, 75\% and 100\%
of all populations fire asynchronously and irregularly at low rate.
Dashed iso-lines confine the area where the firing rates are ordered
in accordance to \prettyref{tab:Comparison-firing-rates}. (C) Population
firing rates of excitatory populations as a function of the relative
inhibitory synaptic strength at fixed background rate ($8\,\mathrm{Hz}$)
(markers as in A).\label{fig:AIness}}

\end{figure}

The low-rate asynchronous irregular (AI) firing regime has been considered
to be the ground state of cortical activity \citep{Amit97}. For the
balanced random network model, the region of stability of the AI state
is bounded by the relative strength of inhibitory synapses ($>4$)
and sufficiently high background rates \citep{Brunel00_183}. \prettyref{fig:AIness}
characterizes the AI state of the layered network. As in the balanced
random network model, the layered network requires sufficient external
input and sufficiently strong inhibitory coupling to enter the AI
state. Increasing the background rate predominantly affects the activity
of L4e, while an increase of the relative inhibitory synaptic strength
decreases mostly the activity of L5e cells. Consequently, the order
of excitatory firing rates (smallest in L2/3 and L6, highest in L5)
is largely preserved except for large relative inhibitory synaptic
strengths (combined with large background rates). 

\subsubsection*{Role of i-specific projections for the stability of the AI state}

\begin{figure}
\begin{centering}
\includegraphics{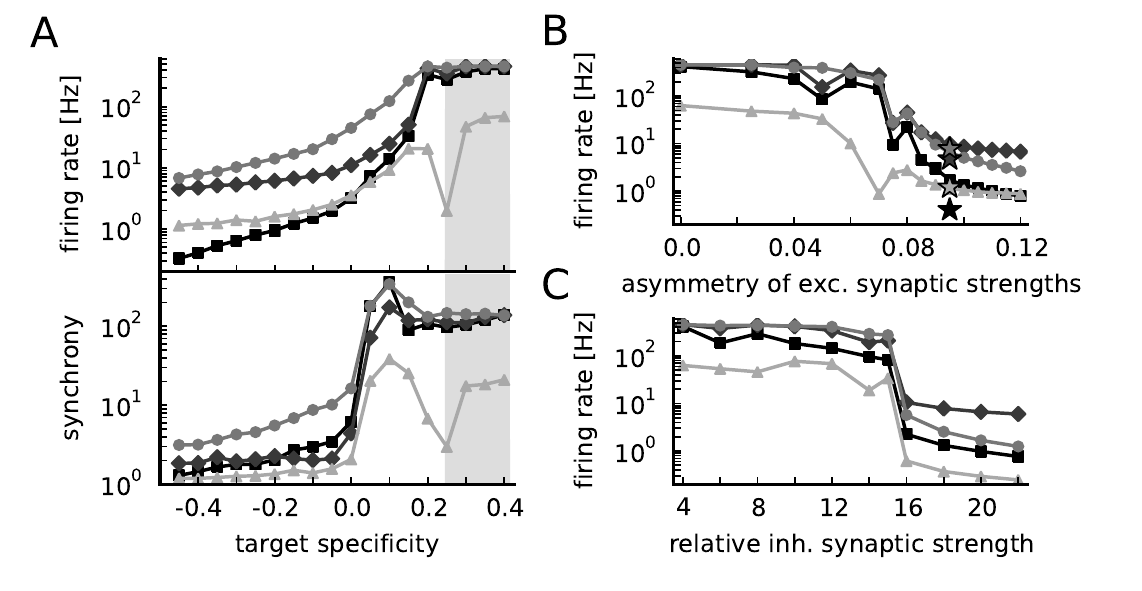}
\par\end{centering}

\caption{Relevance of target specificity for network stability. (A) Population
firing rates (top panel) and synchrony (bottom panel), both in logarithmic
representation, as a function of the target specificity of candidate
projections L2/3e to L4 (negative values twice as large), L2/3e to
L6, L5e to L2/3 and L6e to L4. A target specificity of $0$ reflects
random connectivity, gray-shaded area marks range of target specificity
of other inter-layer connections ($0.33\pm0.08$). (B) Population
firing rates of the model with target specificity of candidate projections
of $+0.4$ as a function of the asymmetry of excitatory to inhibitory
and excitatory to excitatory synaptic strengths (defined as $(w_{\mathrm{ie}}-w_{\mathrm{ee}})/(w_{\mathrm{ie}}+w_{\mathrm{ee}})$,
with fixed mean excitatory synaptic strength). Stars indicate firing
rates of the reference model with target specificity of candidate
projections of $-0.4$. (C) As B, but as a function of the relative
inhibitory synaptic strength. All markers as in \prettyref{fig:AIness}A.\label{fig:tsp-stat}}

\end{figure}

In the following we conduct a series of simulation experiments where
we alter the target type selection of the i-specific projections.
The modifications affect exclusively the target specificity of these
projections (upper four rows in \prettyref{tab:Amendment-candidates-for-tsp}).
Technically, we compile for each parameter set a new connectivity
map algorithmically (see \emph{Materials and Methods}) so that the
experimental data are fully respected. \prettyref{fig:tsp-stat}A
shows that the firing rates and the synchrony of the excitatory populations
first increase exponentially and then superexponentially when the
target specificity of the four candidate projections approach random
connectivity and then the level of other inter-layer connections,
before the spike rates saturate for target specificity values above
$0.2$. The increase in synchronization precedes the firing rate increases.
For L6e, we observe two outliers at $T=0.2$ and $0.25$ where the
activity in this layer corresponds to the low-rate AI state. This
is likely due to the strong within-layer inhibitory feedback in L6
and the high amount of random external inputs to L6e. The general
trend for L6 is nevertheless the same as for the other layers.

The modification of target specificity not only changes the local
microcircuit at the level of specific cell types but also the overall
level of excitation in the network. Therefore, we simulate further
control networks which globally correct for the change in the level
of excitation: we increase for all connections, not only for the i-specific
candidates, the synaptic strength of excitatory to inhibitory connections
and simultaneously reduce the synaptic strength of excitatory to excitatory
connections. \prettyref{fig:tsp-stat}B shows that a sufficiently
large asymmetry of excitatory synaptic strengths with respect to target
cell type counterbalances the overexcitation. However, the natural
order of the excitatory activity levels of the cortical layers (\prettyref{tab:Comparison-firing-rates})
is partly inverted.

This stabilization procedure uses different excitatory synaptic strengths
according to the target cell type and thereby introduces an additional
parameter. Therefore, we also investigate whether the network can
alternatively be stabilized by changing an already existing parameter,
the relative inhibitory synaptic strength. We find that only implausibly
large values ($>15$) lead to a stable low-rate AI state (\prettyref{fig:tsp-stat}C)
and that also in this case the order of firing rates is partly inverted.

In summary, a layered network model that is parametrized equivalent
to the balanced random network model requires the inclusion of i-specific
feedback connections in order to exhibit asynchronous irregular spiking
activity. The alternative stabilization of the network by global changes
of parameters results in a distribution of layer-specific firing rates
in conflict with experimental observations.

\subsection*{Propagation of transient thalamic inputs\phantomsection\addcontentsline{toc}{subsection}{Propagation of transient thalamic inputs}}

\begin{figure}
\begin{centering}
\includegraphics{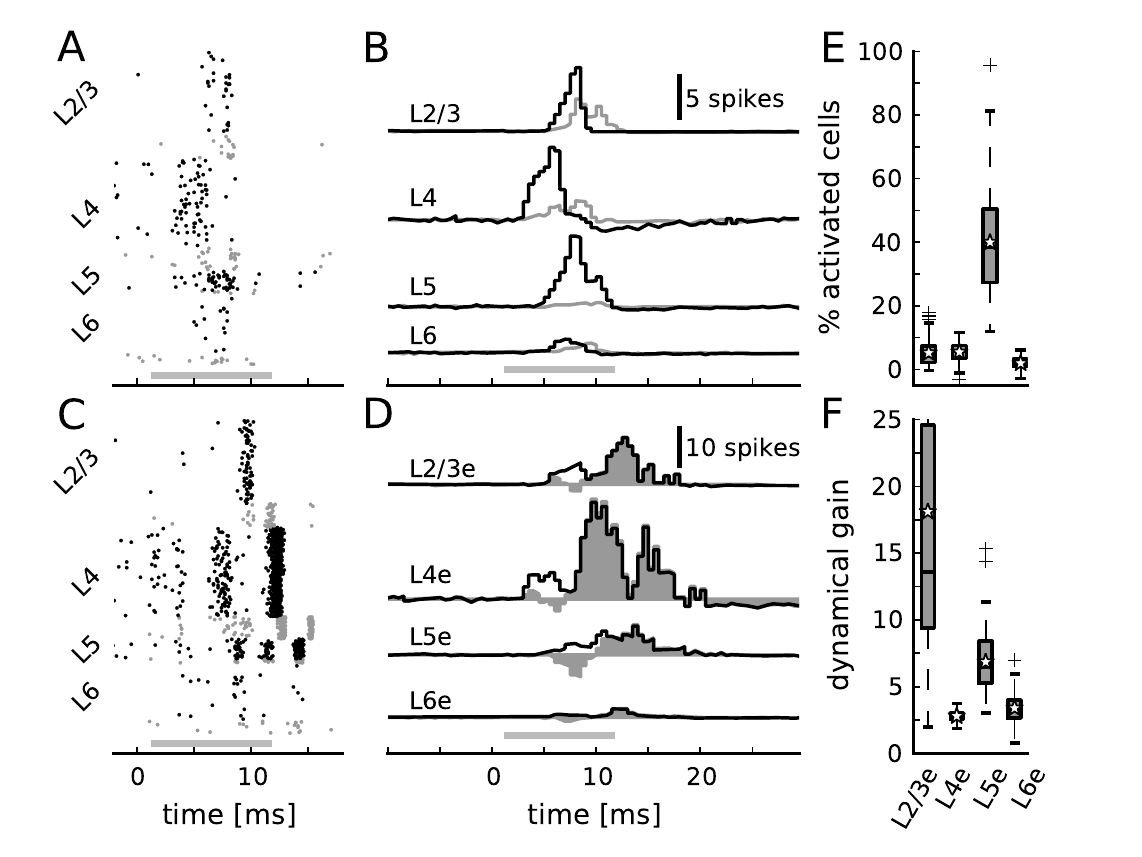}
\par\end{centering}

\caption{Response to transient thalamic input. Thalamic firing rates increase
step-like by $15\,\mathrm{Hz}$ for a duration of $10\,\mathrm{ms}$;
$0\,\mathrm{ms}$ corresponds to the onset of transient input, gray
bars show the arrival of thalamic spikes at cortical neurons taking
the mean delay (for simplicity the same as within the network, $1.5\ms$)
into account. (A) Cell-type specific spiking activity of the network
with i-specific projections. Markers as in \prettyref{fig:stationary-activity}A.
(B) Corresponding cell-type specific population spike counts averaged
over 100 instantiations of network and input (excitatory populations
in black, inhibitory in gray). Spike counts are calculated with a
bin-width of $0.5\,\mathrm{ms}$ with the number of recorded neurons
L3e: 500, L3i: 141, L4e: 529, L4i: 132, L5e: 117, L5i: 25, L6e: 347,
L6i: 71 (numbers correspond to relative population sizes). (C) Spiking
activity of a control network without i-specific projections (target
specificity of candidate projections of $+0.4$) stabilized by an
asymmetry of excitatory synaptic strengths of $0.095$. (D) Corresponding
cell-type specific population spike counts of excitatory populations
averaged over 100 instantiations. Filled area shows the difference
of the data in D and B. (E) Average percentage of cells in the excitatory
populations that are activated by thalamic stimulation in relation
to the ongoing activity (according to the data in B). (F) Dynamical
gain, defined as the firing rate during the stimulus presentation
divided by the spontaneous firing rate, of excitatory populations
(according to data in B).\label{fig:tsp-transient}}

\end{figure}

Confronted with a transient thalamic input the layered network model
responds with a stereotypic propagation of activity through the different
layers. \prettyref{fig:tsp-transient}A shows an exemplary spike raster
of the model after a short-lasting increase of thalamic firing rate.
The cell-type specific activity pattern is consistent over 100 different
network and input instantiations (\prettyref{fig:tsp-transient}B).

\paragraph{Amplitude and timing of cell-type specific responses}

L2/3e, L4e and L5e emit a comparable amount of additional spikes in
response to the stimulus while L6e shows a comparatively sparse response
(\prettyref{fig:tsp-transient}B). In total, only a minority of all
neurons in the network is activated by the thalamic stimulus (\prettyref{fig:tsp-transient}E).
Only in L5e a large fraction of all cells emits additional spikes
in response to the stimulus. In relation to the ongoing activity,
we find that the input layers (L4 and L6) exhibit a similar dynamical
gain which is much lower than the gain of the output layers L5 and
especially L2/3.

The response is initiated in the input layers and then propagates
to layers L2/3 and L5 (see \prettyref{fig:tsp-transient}A, B). The
latency of activation (defined as the maximum of the excitatory spike
count histogram) is shortest in L4 followed by L6 and L5, and finally
L2/3. The early onset of activation of L5, not after but rather synchronously
with L2/3, is in contrast to the expectation according to the classical
notion of the feed-forward loop from L4 to L2/3 to L5 \citep{Gilbert83_217}
but in agreement with the experimental activity data (\citealp{Sakata09_404},
compare also \citealp{Mitzdorf85_37}). The feed-forward connection
from L2/3 to L5 causes the prolonged response in L5 that is reflected
in the second peak in the spike count. L6, being already slightly
excited by the thalamic input, receives feed-forward input from L5
which triggers a sparse response during the ramp-up phase in L5. The
activity is back to baseline in all layers before the last thalamic
spikes arrive at cortex. Deactivation is ordered similar to activation,
starting in L4, followed by L2/3 and L6, and finally L5.

\paragraph{Interplay of excitation and inhibition in the propagation of inputs}

Not only the excitatory populations but also the inhibitory populations
show a distinct activation pattern (see \prettyref{fig:tsp-transient}A,
gray dots, and B, gray lines). Initially, as the interneurons in any
layer receive the same, albeit slightly weaker, feed-forward inputs,
the inhibitory response resembles the excitatory response. However,
the inhibitory populations show, in contrast to the respective excitatory
populations, two pronounced activation peaks in L2/3, L4 and L6. As
shown above, these layers receive i-specific feedback connections.
Apparently, the cell-type specific connectivity structure results
in a complex interplay of excitation and inhibition not explained
by within-layer recurrent inhibition and thereby shapes the propagating
response to a transient thalamic input. Specifically we observe that
the activation of L2/3 entails the increased activity of inhibitory
interneurons in L4, thereby stopping the excitatory activity in this
layer. A similar effect is observed in the next step of the feed-forward
loop between L2/3 and L5. Activation of L2/3 prolongs the excitatory
activity in L5 as stated above. In contrast, the early activation
of L5 results in a sharpening of the response in L2/3 by the additional
activation of L2/3 interneurons. The connection from L6e to L4i adds
to the sharpening of the L4 response and the rather weak i-specific
connection from L2/3 to L6 plays a role in preventing a stronger ramp-up
of excitatory activity in L6.

\paragraph{Role of i-specific feedback connections for the propagation of inputs}

These observations suggest that the i-specific feedback connections
control the duration and the amplitude of the response to thalamic
inputs. To further elucidate their role for the propagation of input-related
activity, we study the response of a network without i-specific feedback
connections but similar spontaneous activity. The small differences
in the firing rates (compare \prettyref{fig:tsp-stat}B) are reflected
in the initial response of the network to the thalamic input (first
bumps in the spike count histograms in \prettyref{fig:activity_flow}D)
which exhibits a steeper increase in L2/3e and L4e and is weaker in
L5e. After this initial phase, the response is drastically different
and shows reverberating activation of the different layers lingering
well beyond the offset of the input. The response is much stronger
and shows oscillatory components by reciprocal activation of the different
layers. The effect stays the same, albeit smaller, for a control network
with the candidate projections having a target specificity value of
$+0.2$ which is well below the range for non-specific inter-layer
connections.

\section*{Discussion\phantomsection\addcontentsline{toc}{section}{Discussion}}

\subsection*{Modeling approach\phantomsection\addcontentsline{toc}{subsection}{Modeling approach}}

The present work extends the balanced random network model \citep{Amit97,Brunel00_183,Vreeswijk96,Vreeswijk98}
to multiple layers with realistic connection probabilities. Despite
their reduced structure the mono-layered models exhibit qualitatively
consistent activity dynamics and the classical analysis of them has
guided our research. Our neuron and synapse model as well as the random
connectivity scheme do not differ qualitatively from the earlier work.
The model size is selected sufficiently large to incorporate the majority
of all local synapses. The network structure, one excitatory and one
inhibitory population in each layer, represents the minimal laminar
extension of the mono-layered models and also the minimal set of cell
types typically distinguished in experiments. Data resolving the connectivity
at a finer scale \citep[e.g.][]{Mercer05_1485,West06_200} are combined
to match the more coarse resolution of our model. Previous multi-layered
models partly use the same approach \citep[e.g.][]{Haeusler09_73}
and partly more detailed cell type classifications \citep[e.g.][]{Traub05_2194,Izhikevich08_3593}.
However, quantitative connectivity data are not yet widely and consistently
available on a finer level of detail so that it remains unclear which
dynamical consequences any further separation of cell types implies.

\subsection*{Integrated connectivity map\phantomsection\addcontentsline{toc}{subsection}{Integrated connectivity map}}

The success of the dynamical analysis in the second part of the study
relies on our finding that the two connectivity maps are consistent
when one considers the differences in methodology. The compiled connectivity
map accounts for these, but it nevertheless mingles data not only
from multiple laboratories but also from different cortical areas
and species. This choice is not exclusively motivated by canonicity
in structure \citep[e.g.][]{Douglas89_480,Nelson02_19,Groh10_826}
and activity (\prettyref{tab:Comparison-firing-rates}), but driven
by the incompleteness of data for a specific area and species. The
density of neurons in our model is based on cat area 17 and the density
of synapses is consistent with cat data \citep{Beaulieu85}. The most
crucial difference between our map and the original data based maps
is the target specific structure which removes the fundamental bias
introduced by undersampling of inhibitory cells (electrophysiology)
or by the application of Peters' rule (anatomy).

The integrated connectivity map is consistent with the most prominent
features of the cortical microcircuit: the recurrence of connections
\citep{Douglas91a} and the feed-forward loop from L4 to L2/3 to L5
to L6 to L4 \citep{Gilbert83_217}. The excitatory sub-circuit is
largely consistent with the recently published excitatory map of the
C2 barrel column of the mouse \citep{Lefort09_301}. In addition,
the circuit exhibits a distinct feedback structure with projections
targeting predominantly interneurons from L2/3e to L4 and from L5e
to L2/3. \citet{Thomson02_1781,Thomson02} recognized the potential
of selective feedback projections for stabilizing the activity and
increasing the sensitivity for time-dependent signaling, but were
not able to test this hypothesis. Furthermore, reports on the projections
from L6e to L4 and the rather weak projection from L2/3e to L6 indicate
specific selection of inhibitory target cells. Our selection of the
projection from L6e to L4 is based on the electron microscopy study
by \citet{McGuire84_3021}, providing sparse and partial data. In
their discussion, \citet{Ahmed94} present an alternative interpretation
of these data proposing that many synapses originally assigned to
inhibitory targets are potentially on excitatory cells. Still they
acknowledge that the relative number of synapses targeting interneurons
is very high in comparison to other excitatory projections. The ongoing
discussion on this projection has recently been reviewed by \citet{Thomson10_13}.

Here, we quantified the dynamical relevance of these i-specific connections
for spontaneous and evoked activity. Especially the two feedback projections
(L2/3e to L4 and L5e to L2/3) are crucial for stability and reliable
input propagation because of the relatively small structural and dynamical
impact of L6 on other layers. Extending the idea of i-specific feedback,
it is a conceivable but untested hypothesis that L6e (especially cortico-thalamic
cells) targets predominantly interneurons in L5, too.

\subsection*{Model parametrization\phantomsection\addcontentsline{toc}{subsection}{Model parametrization}}

Only a subset of the required model parameters are known experimentally.
Here, we focus on cell-type specific structural parameters: the numbers
of neurons, the numbers of external inputs and the connection probabilities
between neurons. Other parameters like neuronal parameters, synaptic
strengths and delays are selected independent of the cell type. This
choice enables us to expose the dynamical consequences of the structure
of the local microcircuit unaffected by additional cell-type specificity
\citep[e.g.][]{Bremaud07_14134,Lefort09_301}. The consistency of
the activity in our model with the experimental data suggests that
the (static) connectivity structure plays a dominant role in shaping
the neuronal activity and that it is not required to model complex
neuronal features such as morphology to reproduce the experimental
findings discussed here. In future, modeling of more complex neuronal
dynamics will eventually reveal where spatially extended neuron models
are required to understand the experimentally observed phenomena.

An exception to the cell-type independent parametrization is the increased
synaptic strength for the connection from L4e to L2/3e. Although this
change affects the ongoing activity only marginally (supplementary
Fig. 19), it is important for a successful transmission of activity
from L4 to L2/3 following thalamic stimulation. The modification is
motivated by the large discrepancies of the L4i to L2/3 projection
in the anatomical and physiological maps and the difference in the
relative convergence of excitatory inputs from L2/3 and L4 to L2/3
pyramids between our model and \citet{Feldmeyer06_583}. To ultimately
resolve this issue, it might be necessary to incorporate additional
specificity \citep{Yoshimura05_868,Sarid07_16353,Fares09_16463} and,
despite a plethora of studies on the L4 to L2/3 connections, potentially
additional experiments especially regarding the inhibitory projection.

\subsection*{Cell-type specific activity\phantomsection\addcontentsline{toc}{subsection}{Cell-type specific activity}}

\begin{figure}
\begin{centering}
\includegraphics{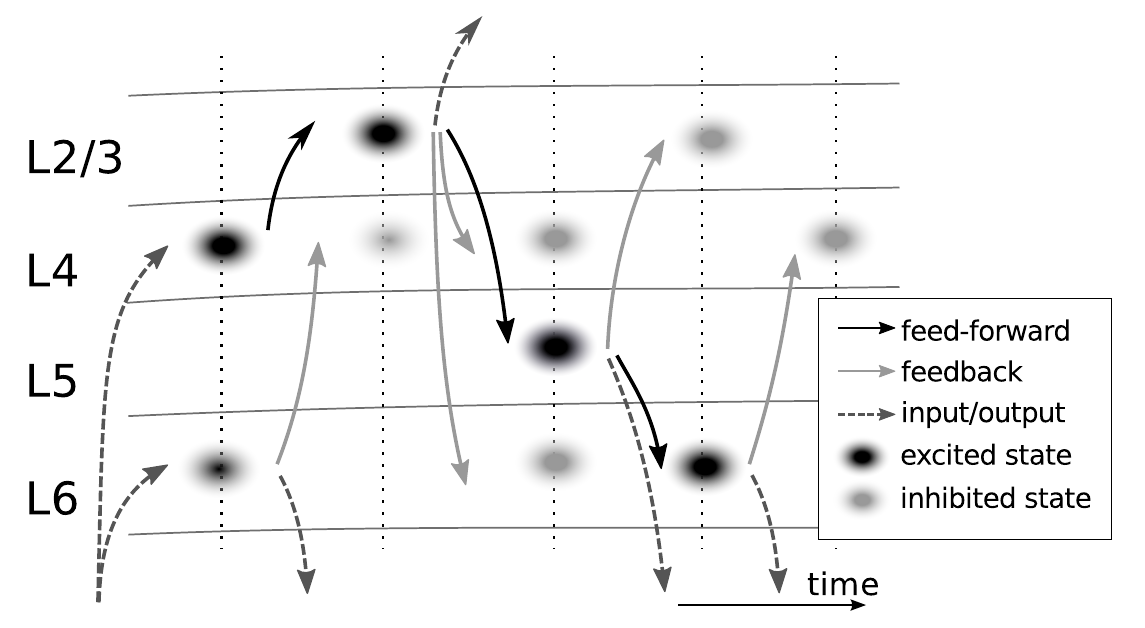}
\par\end{centering}

\caption{Sketch of the flow of activity following transient thalamic input.
Black and gray ellipses represent excited and inhibited activity states,
respectively. Dark gray dashed arrows indicate input and output of
the local network. The black arrows represent the feed-forward loop
projections L4 to L2/3 to L5 to L6. The gray arrows correspond to
the activation of the candidate i-specific connections (L2/3e to L4
and to L6, L5e to L2/3 and L6e to L4).\label{fig:activity_flow}}

\end{figure}

The application of the integrated connectivity map and the parametrization
according to balanced random network models \citep{Brunel00_183}
results in asynchronous irregular activity without specific tuning.
This activity state is stable over a wide range of parameters regarding
e.g. external inputs, synaptic strengths and delays.

The model predicts cell-type specific firing rates in agreement with
data from awake animals (\prettyref{tab:Comparison-firing-rates}).
In particular, the connectivity map captures the low excitatory firing
rates in L2/3 and L6 quantitatively although mono-layered models hardly
show stable activity at these low levels of activity \citep{Sussillo07_4079}.
The particular ordering of firing rates depends, however, on the inclusion
of i-specific feedback connections. The lacking consistency of the
firing rates in previous models is presumably due to the incompleteness
of the previously used connectivity maps.

Confronted with a transient thalamic input, the model shows a concise
propagation of activity from the input layers to the output layers
(compare \citealp{Miller96}). The propagation pattern (\prettyref{fig:activity_flow})
is shaped by the interaction of excitation and inhibition between
the different layers and promotes a temporal neural code \citep{Thomson02}.

\subsection*{Relation of structure and activity\phantomsection\addcontentsline{toc}{subsection}{Relation of structure and activity}}

\begin{figure}
\centering{}\includegraphics{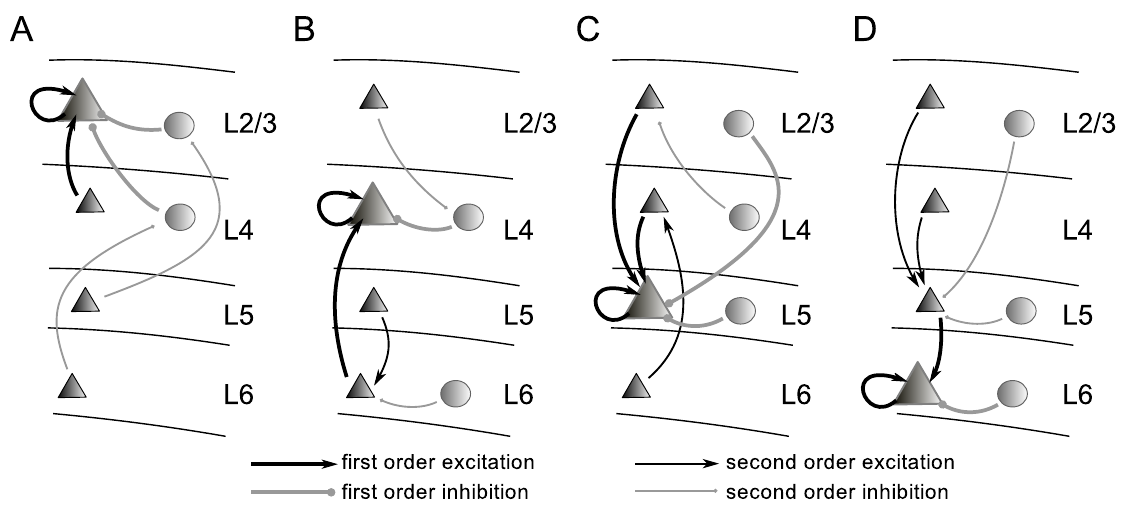}\caption{Differences in input structure for the excitatory cell types L2/3e
(A), L4e (B), L5e (C) and L6e (D) (large triangles). The illustrations
show the strongest pathways of direct (first-order, thick arrows)
and indirect (second-order, thin arrows) excitation (black) and inhibition
(gray) of a given population (see \emph{Supplemental Material} for
further details). Triangles represent excitatory and circles inhibitory
populations.\label{fig:inputs_to_Eneurons}}

\end{figure}
The differences in the cell-type specific input structure (\prettyref{fig:inputs_to_Eneurons})
shed light on the mechanisms underlying the observed activity features.
For example, the low firing rates of L2/3e and L6e neurons have different
structural origins: L2/3e effectively integrates, next to the excitatory
inputs from L2/3e and L4e, inhibitory inputs from all layers. In contrast,
L6e interacts largely with the recurrent within-L6 network; other
inputs predominantly pass through L5e and rather modulate L6e activity.
L4e is also dominated by within-layer connectivity and the i-specific
inputs from L2/3 and L6 modulate its activity and temporally structure
the response to transient stimuli. L5 consists of relatively few neurons
and correspondingly forms rather few recurrent within-layer inputs.
Furthermore, L5e integrates the highest number of first-order inputs,
but does not receive i-specific projections. As a result, L5e reacts
with a wide range of firing rates to changes in the external input,
especially in comparison to the other layers. This, together with
the control that L5 exerts on L2/3 by the i-specific feedback, puts
it in a special position to integrate and amplify information. Inhibitory
cells receive predominantly excitatory inter-layer inputs (supplementary
Fig. 15) and therefore exhibit elevated firing rates.

\subsection*{Conclusion\phantomsection\addcontentsline{toc}{subsection}{Conclusion}}

The connectivity structure of the local cortical network shapes the
cell-type specific activity and defines functional roles of the different
layers. Any fine-scale connectivity structure \citep[e.g.][]{Song05_0507,Yoshimura05_868,Kampa06_1472}
faces the constraints imposed by the connectivity implemented in this
model. The presented framework can be continuously refined as new
data become available and extends the available mathematical methods
to infer synaptic connectivity from neuronal morphology. Based on
the currently available data it reproduces prominent activity features,
suggesting that these arise predominantly from the network structure,
not single cell properties. The results predict distinct activity
patterns of interneurons and highlight the need to uncover the target
specificity of projections in future experiments.

\paragraph*{Acknowledgements:}

We are grateful to Kathleen S. Rockland and Dirk Feldmeyer for fruitful
discussions about the structure of the local cortical network, and
to our colleagues in the NEST Initiative for continued support. We
also acknowledge Ad Aertsen for his contributions in the initial phase
of the project. Partially supported by the Helmholtz Alliance on Systems
Biology, EU Grant 15879 (FACETS), EU Grant 269921 (BrainScaleS), DIP
F1.2, BMBF Grant 01GQ0420 to BCCN Freiburg, and the Next-Generation
Supercomputer Project of MEXT (Japan).

\newpage

\bibliographystyle{neuralcomput_natbib}
\bibliography{brain}

\newpage

\section*{Supplemental Material}

\subsection*{Connection probabilities}

\subsubsection*{Derivation of Eq. (1)}

The connection probability $C_{N^{\mathrm{pre}},N^{\mathrm{post}}}(K)$
of two neurons from populations of sizes $N^{\mathrm{pre}}$ and $N^{\mathrm{post}}$
which are randomly connected (with uniform probability) by $K$ synapses
can be derived as follows: the probability that a particular pair
of neurons is connected is equal to one minus the probability that
it is not connected, i.e. \begin{eqnarray*}
C_{N^{\mathrm{pre}},N^{\mathrm{post}}}(K) & = & 1-q_{N^{\mathrm{pre}},N^{\mathrm{post}}}(K).\end{eqnarray*}
Since the $K$ synapses are created independently, $q_{N^{\mathrm{pre}},N^{\mathrm{post}}}(K)$
is the probability of not being connected by any of the synapses \begin{eqnarray*}
q_{N^{\mathrm{pre}},N^{\mathrm{post}}}(K) & = & \prod_{i=1}^{K}q_{N^{\mathrm{pre}},N^{\mathrm{post}}}(1)=[q_{N^{\mathrm{pre}},N^{\mathrm{post}}}(1)]^{K}.\end{eqnarray*}
The probability of not being connected by a particular synapse is
one minus the probability of being so \begin{eqnarray*}
q_{N^{\mathrm{pre}},N^{\mathrm{post}}}(1) & = & 1-C_{N^{\mathrm{pre}},N^{\mathrm{post}}}(1)\\
 & = & 1-1/N^{\mathrm{pre}}N^{\mathrm{post}},\end{eqnarray*}
hence \begin{eqnarray*}
C_{N^{\mathrm{pre}},N^{\mathrm{post}}}(K) & = & 1-\left(1-\frac{1}{N^{\mathrm{pre}}N^{\mathrm{post}}}\right)^{K}.\end{eqnarray*}

\subsubsection*{Taylor-series expansion of Eq. (1)}

The expression for the connection probability can be expanded in a
Taylor series using the binomial series \begin{eqnarray*}
(1+x)^{K} & = & 1+Kx+\frac{K(K-1)}{2}x^{2}+...\end{eqnarray*}
Although conversion is guaranteed for $x<1$, we require $K^{2}x^{2}\to0$
in order to write $(1+x)^{K}=1+Kx$. With $x=-1/N^{\mathrm{pre}}N^{\mathrm{post}}$
we find the approximation\begin{eqnarray*}
C_{N^{\mathrm{pre}},N^{\mathrm{post}}}(K) & = & 1-\left(1-\frac{1}{N^{\mathrm{pre}}N^{\mathrm{post}}}\right)^{K}\\
 & = & 1-(1+x)^{K}\\
 & = & 1-(1+Kx)\\
 & = & -Kx\\
 & = & \frac{K}{N^{\mathrm{pre}}N^{\mathrm{post}}}\end{eqnarray*}
valid for small $K/N^{\mathrm{pre}}N^{\mathrm{post}}$.

\subsubsection*{Errors of experimental connection probabilities}

A lower bound for the statistical error of $C_{\mathrm{a}}$ results
from the experimental standard deviations of the neuron count by \citet{Beaulieu83}
and the consideration of error propagation.

The physiological search for connected neurons corresponds to a random
sampling test, so that the statistical error of the number of connected
neuron pairs $\delta(RQ)$ is given by the standard deviation of
the binomial distribution. Therefore, the statistical error of the
connection probability is $\delta C_{\mathrm{p}}=\sqrt{C_{\mathrm{p}}(1-C_{\mathrm{p}})/Q}$
(given $Q>0$). For untested connections (i. e. $\sum_{j}Q_{j}=0$),
we set $\delta C_{\mathrm{p}}=0.01$.

Both estimates are minimal statistical errors. The estimate for the
anatomical data ignores errors in the estimation of the number of
synapses as well as errors due to the limited number of reconstructed
axons and dendrites. Both error estimates do not account for any potential
systematic error e.g. due to sampling biases. The minimal statistical
errors are sufficient to show that recurrence strength and loop strength
of the two connectivity maps (see Fig. 2) are statistically indistinguishable.

\subsection*{Gaussian lateral connectivity model}

The anatomical and physiological mean connection probabilities $C_{\mathrm{a/p}}$
correspond to the integration of the lateral connectivity profile
$C(r)$ over the corresponding sampling radius $r_{\mathrm{a/p}}$,
i.e. $C_{\mathrm{a}/\p}=1/(\pi r_{\a/\p}^{2})\int_{0}^{r_{\mathrm{a}/\p}}\int_{0}^{2\pi}C(r)rdrd\varphi$,
yielding the two equations \begin{eqnarray}
C_{\mathrm{a}} & = & 2\pi C_{0}\sigma^{2}/(\pi r_{\a}^{2})[1-\exp(-r_{\a}^{2}/2\sigma^{2})]\label{eq:Ca-integration-res}\\
C_{\p} & = & 2\pi C_{0}\sigma^{2}/(\pi r_{\p}^{2})[1-\exp(-r_{\p}^{2}/2\sigma^{2})].\label{eq:Cp-integration-res}\end{eqnarray}
Assuming that the parameters of the lateral connectivity profile $C_{0}$
and $\sigma$ are identical in both equations, we find\begin{eqnarray*}
\frac{\pi r_{\a}^{2}C_{\mathrm{a}}}{1-\exp(-r_{\mathrm{a}}^{2}/2\sigma^{2})} & = & \frac{\pi r_{\mathrm{p}}^{2}C_{\mathrm{p}}}{1-\exp(-r_{\p}^{2}/2\sigma^{2})}.\end{eqnarray*}
This equation can be solved numerically for $\sigma$. For $r_{\mathrm{a}}\gg\sigma$
we have $\exp(-r_{\mathrm{a}}^{2}/2\sigma^{2})=0$ and therefore\begin{eqnarray*}
\exp(-r_{\p}^{2}/2\sigma^{2}) & = & 1-\frac{\pi r_{\mathrm{p}}^{2}C_{\mathrm{p}}}{\pi r_{\a}^{2}C_{\mathrm{a}}}\end{eqnarray*}
which can be solved for $\sigma$:\begin{eqnarray*}
\sigma & = & r_{\mathrm{p}}\left[-2\ln\left(1-\frac{\pi r_{\mathrm{p}}^{2}C_{\mathrm{p}}}{\pi r_{\a}^{2}C_{\mathrm{a}}}\right)\right]^{-1/2}.\end{eqnarray*}
$C_{0}$ can now be found by solving \prettyref{eq:Ca-integration-res}
(or analogously \prettyref{eq:Cp-integration-res}):

\begin{eqnarray*}
C_{0} & = & \frac{\pi r_{\mathrm{a}}^{2}C_{\mathrm{a}}}{2\pi\sigma^{2}}\left[1-\exp\left(-\frac{r_{\mathrm{a}}^{2}}{2\sigma^{2}}\right)\right]^{-1}.\end{eqnarray*}
In practice, we are interested in the limit $r_{\mathrm{a}}\to\infty$,
arriving at eq. (5) for $\sigma$ and, using $\lim_{r_{\mathrm{a}}\to\infty}\exp(-r_{\mathrm{a}}^{2}/2\sigma^{2})=0$,
eq. (6) for $C_{0}$.

The resulting parameters depend on the sampling radius of the physiological
experiments, i.e. the lateral distance of the somata of cells in the
paired recordings. For the raw data set, this value is provided ($100\mum$,
\citealp{Thomson02}). The modified physiological map, however, combines
different experiments with potentially different sampling radii. Here,
we find that increasing the sampling radius increases the zero-distance
connection probability monotonically but decreases the lateral spread.
Nevertheless, the effect is small: the estimates of both parameters
change by less than 8\% when altering the sampling radius from $50$
to $150\mum$.

\subsection*{Scaling factor}

The scaling factor $\zeta$ is used to compare measurements of connection
probabilities of individual connections. In order to first remove
global differences that arise from the differences in the lateral
sampling, we scale the connection probabilities $C_{\mathrm{a/p}}$
of a connectivity map such that the new mean of the map corresponds
to a previously determined model connectivity $C_{\mathrm{m}}$: $C'_{\mathrm{a/p}}\leftarrow C_{\mathrm{a/p}}C_{\mathrm{m}}/\bar{C}_{\mathrm{a/p}}$,
where $\bar{C}_{\mathrm{a/p}}$ denotes the global mean of the original
connectivity map. The ratio of the scaled individual connection probabilities
is $1$ if two maps differ with respect to the lateral sampling but
are otherwise in perfect agreement. In order to exclusively measure
the quality of the agreement, we remove the information which connectivity
map provides the larger estimate by calculating the ratio of the larger
of the two estimates divided by the smaller one\begin{eqnarray*}
\zeta & = & \frac{\max(C'_{\mathrm{a}},C'_{\mathrm{p}})}{\min(C'_{\mathrm{a}},C'_{\mathrm{p}})}\qquad\mathrm{if}\, C'_{\mathrm{a/p}}>0,\end{eqnarray*}
restricting the measure to $\zeta>1$. Instead of scaling to a previously
determined model connectivity, it is equivalent to compare the normalized
connection probabilities\begin{eqnarray*}
\zeta & = & \frac{\max(C'_{\mathrm{a}},C'_{\mathrm{p}})}{\min(C'_{\mathrm{a}},C'_{\mathrm{p}})}\\
 & = & \frac{\max(C_{\mathrm{a}}C_{\mathrm{m}}/\bar{C}_{\mathrm{a}},C_{\mathrm{p}}C_{\mathrm{m}}/\bar{C}_{\mathrm{p}})}{\min(C_{\mathrm{a}}C_{\mathrm{m}}/\bar{C}_{\mathrm{a}},C_{\mathrm{p}}C_{\mathrm{m}}/\bar{C}_{\mathrm{p}})}\\
 & = & \frac{\max(C_{\mathrm{a}}/\bar{C}_{\mathrm{a}},C_{\mathrm{p}}/\bar{C}_{\mathrm{p}})}{\min(C_{\mathrm{a}}/\bar{C}_{\mathrm{a}},C_{\mathrm{p}}/\bar{C}_{\mathrm{p}})}.\end{eqnarray*}
This can be seen by substituting the mathematical definitons of $\max(x,y)$
and $\min(x,y)$, $|x+y|\pm|x-y|$.

\subsection*{Consistent modification of target specificity: exact expression for
redistribution of synapses}

The redistribution of synapses is a method to modify the connection
probabilities forming a projection while conserving the total number
of synapses. The connection probability is a non-linear function of
the number of synapses (see Eq. (1)) and therefore the proportion
of synapses that contacts excitatory targets $\Delta$ depends on
the number of neurons in the presynaptic and the two postsynaptic
populations ($N^{\mathrm{pre}}$, $N^{\mathrm{post=e}}$ and $N^{\mathrm{post=i}}$)
as well as on the total number of synapses $K$ and the desired target
specificity $T$. In \emph{Materials and Methods}, we provide the
solution for $\Delta$ assuming the linear approximation $C=K/(N^{\mathrm{pre}}N^{\mathrm{post}})$.
The exact value is found by numerically solving 

\begin{eqnarray}
2T & = & \left(1-\frac{1}{N^{\mathrm{post=i}}N^{\mathrm{pre}}}\right)^{(1-\Delta)K}(1+T)\nonumber \\
 &  & -\hspace{3pt}\left(1-\frac{1}{N^{\mathrm{post=e}}N^{\mathrm{pre}}}\right)^{\Delta K}(1-T)\label{eq:redistribution_exact}\end{eqnarray}
for $\Delta$. In practice, we use this variant in order to prevent
inaccuracies due to the approximation if $K/(N^{\mathrm{pre}}N^{\mathrm{post}})$
is not sufficiently small.

\begin{figure}
\begin{centering}
\hspace*{-1cm}\includegraphics{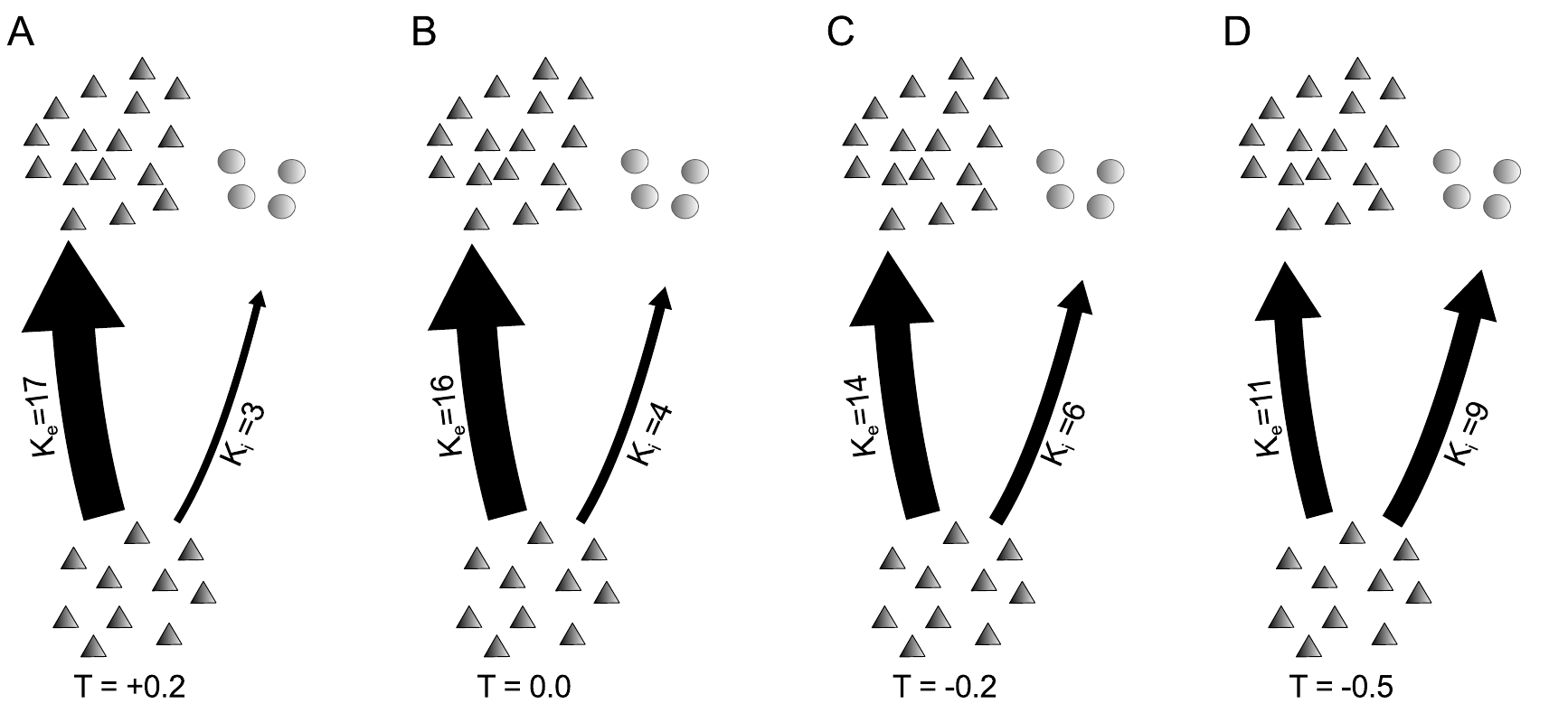}
\par\end{centering}

\caption{Redistribution of synapses with conserved total number of synapses
according to eq. \eqref{eq:redistribution_exact}. The cartoons depict
a situation with $N^{\mathrm{pre}}=10$ presynaptic neurons (bottom
triangles), $N^{\mathrm{post=e}}=16$ excitatory (top triangles) and
$N^{\mathrm{post=i}}=4$ inhibitory (circles) postsynaptic neurons
and $K=20$ synapses forming the projection of the presynaptic population
to the postsynaptic layer. The arrow thickness corresponds to the
indicated number of synapses that select excitatory ($K_{\mathrm{e}}$)
and inhibitory ($K_{\mathrm{i}}$) targets. (A) Due to the larger
dendritic length of the postsynaptic excitatory neurons, the application
of Peters' rule results in a preferential targeting of excitatory
neurons with a target specificity of at least $T=+0.2$ (Fig. 4).
(B) Random selection of postsynaptic neurons as it is typically applied
in mono-layered network models (e.g. \citet{Brunel00}). (C) and (D)
show preferental selection of inhibitory targets with a target specificity
of $T=-0.2$ and $T=-0.5$, respectively.}

\end{figure}

\subsection*{Selection of additional candidates for specific target type selection}

Electrophysiological recordings \citep{Thomson02_936} demonstrated
that the projection from L2/3e to L4 exclusively targets interneurons.
We are not aware of further studies that investigate the inputs from
L2/3 to L4 inhibitory cells. However, no further study in visual or
somatosensory areas found connections from L2/3e to L4e, but connections
in the inverse direction were reported from L4e to L2/3e (see Table
5). Only in auditory cortex, projections from L2/3e to L4e have been
reported to pyramids and star pyramids \citep{Barbour08_11174}.

The projection from L5e to L2/3 has a large scaling factor (Fig. 2)
and evidence exists from several studies that this projection targets
specifically inhibitory neurons in L2/3: Strongest evidence comes
from a photostimulation study \citep{Dantzker00} reporting that L2/3e
cells receive few inputs from L5e while a subset of L2/3 inhibitory
neurons receives the majority of its inputs from this layer. For the
inputs from deep L5, only the input to this type of interneurons is
significant at all. Yet, no paired recordings study reports measurements
of the L5e to L2/3i connection, but several studies report an asymmetry
of the feed-forward L2/3e to L5e and the inverse L5e to L2/3e connection
(\citealp{Thomson98_669,Thomson02_936}: ten-fold higher feed-forward
connectivity, \citealp{Lefort09_301}: 3.6-fold higher feed-forward
connectivity). In contrast, the anatomical estimate provides similar
probabilities for both connections and, in addition to \citet{Binzegger04}
which is based on a single reconstructed L5e axon projecting to L2/3,
several studies found L5e cells predominantly projecting to L2/3 (e.g.
\citealp{Martin84,Stepanyants07}). These discrepancies between physiological
and anatomical estimates can partly be related to undersampling of
cells projecting to L2/3 in physiological recordings and also to stronger
cutting of ascending axons in slice experiments: \citet{Stepanyants09_3555}
estimated that only 60\% of all potential connections can be expected
to be intact for the excitatory L5 to L2/3 connection in paired recordings
in slices. A refinement of our model that takes into account multiple
excitatory L5 populations (e.g. L5A and L5B, see \citealp{Schubert06_223})
can potentially resolve some of the connectivity observed on a finer
scale. Taken together, this evidence qualifies the L5e to L2/3 projection
as an i-specific candidate projection.

\citet{Zarrinpar06_1751} investigated the layer-specific input to
L6 pyramids and interneurons in photostimulation experiments in rat
primary visual cortex. Although the inputs from L2/3 to L6 are found
to be sparse, they are significantly preferring interneuronal targets:
in only very few cases (less than 10\%) significant excitatory inputs
from L2/3 to L6 pyramids are found, but in around 70\% of all trials
to interneurons (see their figure 5).

\citet{McGuire84_3021} find in an electron microscopy (EM) study
that asymmetric (excitatory) synapses originating in L6 preferentially
target the shafts of dendrites in L4. We include this projection as
a candidate for specific target type selection inspite of the few
samples underlying this study and the indirect assessment of connectivity
without explicit identification of the postsynaptic cell. This interpretation
of the EM data is partly supported by a similar case in the literature:
\citet{White87_13} and \citet{Elhanany90_43} applied EM to study
the connections of L6 cortico-cortical (CC) and cortico-thalamic (CT)
cells and found that CT cells target predominantly dendritic shafts
(as the L6e to L4 projection in\citet{McGuire84_3021}) while CC cells
target spines as typical for excitatory cells \citep{Braitenberg98}.
Later, paired recordings \citep{Mercer05_1485,West06_200} showed
that this ultrastructural feature indeed reflects specific target
type selection of interneurons. Since primarily CT cells project to
L4 \citep{Briggs10_3} it is possible that these cells show the same
target type selection in this layer as in L6.  Although the evidence
is rather sparse (see also \emph{Discussion}), we include also this
projection in the list of i-specific candidates.

For the simulations, our reference parametrization assumes that the
target specificity of the i-specific projections is $-0.8$ for the
projection from L2/3e to L4 and $-0.4$ for the three other candidate
projections.

\subsection*{Algorithmic compilation of integrated connectivity map}

\begin{enumerate}
\item Input: 

\begin{enumerate}
\item $C_{\mathrm{a}}=1-\left(1-\frac{1}{N^{\mathrm{pre}}N^{\mathrm{post}}}\right)^{K}$
\item $C_{\mathrm{p}}=\sum_{i}R_{i}Q_{i}/(\sum_{j}Q_{j})$
\item define $r_{\mathrm{m}}$ (here $1\mm^{2}$)
\item define amendment candidates for target specificity $T$ (Table 2)
\end{enumerate}
\item Scale connection matrices

\begin{enumerate}
\item $\sigma=r_{\mathrm{p}}\left[-2\ln\left(1-\frac{\tilde{C}_{\mathrm{p}}}{\tilde{C}_{\mathrm{a}}}\right)\right]^{-1/2}$
\item $C_{0}=\frac{\tilde{C}_{\mathrm{a}}}{2\pi\sigma^{2}}$
\item $\bar{C}_{\mathrm{m}}=\frac{2}{r_{\mathrm{m}}^{2}}C_{0}\sigma^{2}[1-\exp(-r_{\mathrm{m}}^{2}/(2\sigma^{2}))]$
\item $C{}_{\mathrm{a/p}}\leftarrow C_{\mathrm{a/p}}\bar{C}_{\mathrm{m}}/\bar{C}_{\mathrm{a/p}}$
\end{enumerate}
\item Redistribute synapses of anatomical map

\begin{enumerate}
\item solve \prettyref{eq:redistribution_exact} for $\Delta$
\item $C_{\mathrm{a}}^{\mathrm{post=e}}\leftarrow1-\left(1-\frac{1}{N^{\mathrm{pre}}N^{\mathrm{post}}}\right)^{\Delta K}$
(inhibitory connections analogously)
\end{enumerate}
\item Correct target specificity of physiological map

\begin{enumerate}
\item $C_{\mathrm{p}}^{\mathrm{post=i(e)}}\leftarrow\left(\frac{1-T}{1+T}\right)^{+(-)1}C_{\mathrm{p}}^{\mathrm{post=e(i)}}$
\end{enumerate}
\item $C_{\mathrm{m}}=\frac{1}{2}(C{}_{\mathrm{a}}+C{}_{\mathrm{p}})$
\end{enumerate}
\begin{figure}
\begin{centering}
\includegraphics{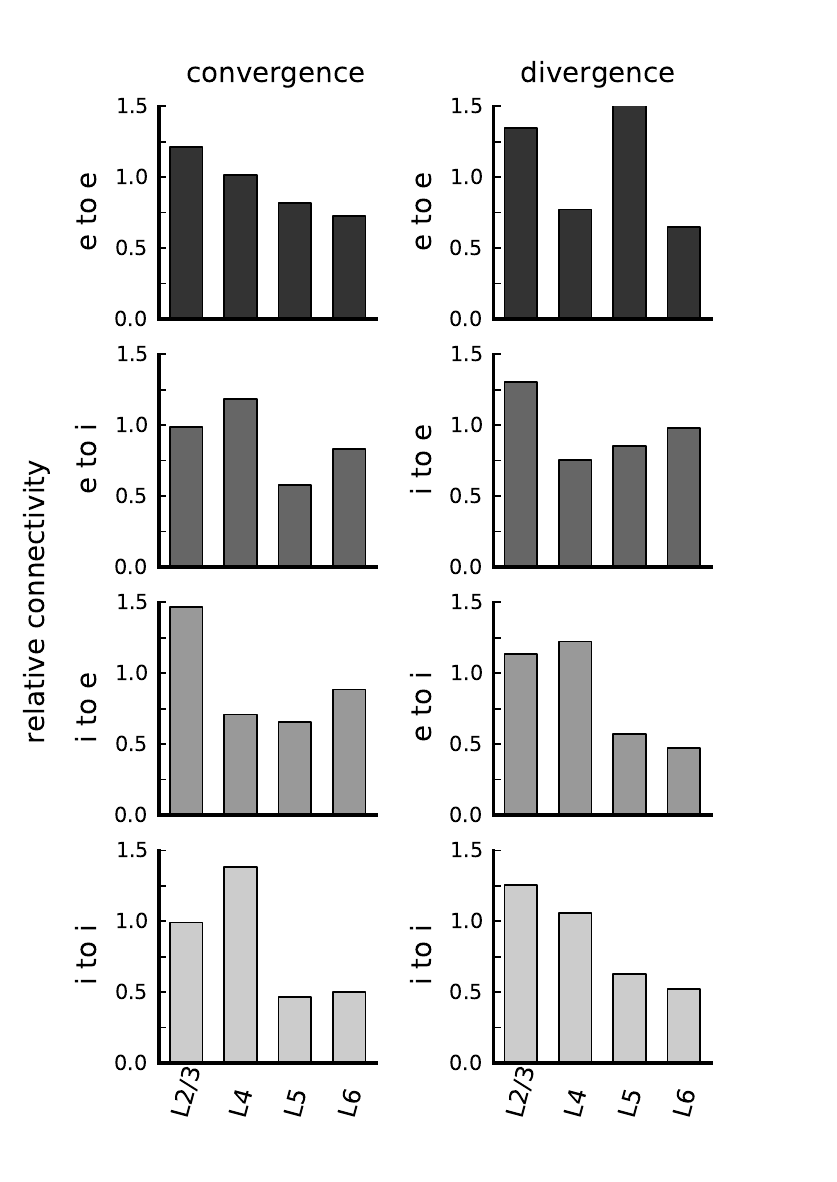}
\par\end{centering}

\caption{Relative connectivity of the four connection types. The relative convergent
(left column) connectivity is the ratio of the relative averaged convergence
(weighted according to neuronal densities of postsynaptic cell types)
and the relative neuronal densities of the presynaptic cell types.
The relative divergent (right column) connectivity is defined analogously
for divergence. The relative connectivity can be used to determine,
which layer plays the dominant role in the local network: If the overall
effect of a neuron would be independent of its layer, any neuron in
the network, on average, sampled randomly from all other neurons and
the layer specificity of the average convergence and divergence corresponded
to the distribution of the number of neurons (relative connectivity
of $1$). We find that neurons do not sample randomly from all other
neurons in the network, for different connection types, different
layers play the dominant role: the highest number of excitatory inputs
relative to neuron numbers (to excitatory and inhibitory targets)
originates from L2/3, inhibitory inputs (again to both target types)
from L4. Fewest inputs relative to neuron numbers are provided by
L6 (e to e) and L5 (all other connection types). For outputs, the
situation is different: the highest number of outputs relative to
neuron numbers are projected to L5 (e to e), to L4 (e to i) and to
L2/3 (i to e/i). Fewest outputs relative to neuron numbers are projected
to L4 (i to e) and to L6 (all other types).\label{fig:Specificity-of-connections.}}

\end{figure}

\subsection*{Gray-matter long-range inputs}

In order to estimate the layer-specific number of external gray-matter
inputs $K_{\mathrm{ext}}^{\mathrm{gm}}$, we first define the relative
number of external gray-matter connections as $\gamma_{\mathrm{ext}}=K_{\mathrm{ext}}^{\mathrm{gm}}/K_{\mathrm{total}}^{\mathrm{gm}}$,
where the total number of gray-matter connections can be divided into
local and external connections $K_{\mathrm{total}}^{\mathrm{gm}}=K_{\mathrm{local}}+K_{\mathrm{ext}}^{\mathrm{gm}}$.
Here, we assume that the relative number of external gray-matter connections
$\gamma_{\mathrm{ext}}$ corresponds to the weighted and normalized
number of bouton clusters located outside of the local cortical network
as measured by \citet{Binzegger07_12242}. The weighted and normalized
number of non-local bouton clusters is defined as follows: we count
the number of bouton clusters $\xi_{\rho}$ with a horizontal displacement
greater than $0.56\,\mathrm{mm}$ (i.e. outside of the local cortical
volume below $1\,\mathrm{mm}^{2}$ of cortical surface) in \citet{Binzegger07_12242},
their figure 11, according to the rank $\rho\in\{1,2,3\}$ (where
rank $3$ collapses rank $>2$). The relative number of external bouton
clusters is calculated by dividing $\xi_{\rho}$ by the total number
of clusters $\Xi_{\mathrm{total}}$ that can be extracted from \citet{Binzegger07_12242},
their figure 3B. In order to assess the relative proportion of synapses
formed by neurons outside of the local cortical network, we multiply
the relative number of external bouton clusters with the cluster weights
$\omega_{\rho}$ that can be extracted from \citet{Binzegger07_12242},
their figures 4C and 4D for rank $\rho=1$ and $2$. As an estimate
for rank $3$ we use the remaining weight $\omega_{3}=1-(\omega_{1}+\omega_{2})$.
Finally, we sum over all ranks and obtain with \begin{eqnarray*}
\gamma_{\mathrm{ext}} & = & \sum_{\rho=1}^{3}\xi_{\rho}\omega_{\rho}/\Xi_{\mathrm{total}}\end{eqnarray*}
 for the number of external gray-matter inputs:\begin{eqnarray*}
K_{\mathrm{ext}}^{\mathrm{gm}} & = & K_{\mathrm{local}}\frac{\gamma_{\mathrm{ext}}}{1-\gamma_{\mathrm{ext}}}\\
 & = & K_{\mathrm{local}}\frac{\sum_{\rho=1}^{3}\xi_{\rho}\omega_{\rho}}{\Xi_{\mathrm{total}}-\sum_{\rho=1}^{3}\xi_{\rho}\omega_{\rho}}.\end{eqnarray*}
The total number of layer-specific external inputs $K_{\mathrm{ext}}$
then is the sum of the external gray-matter inputs $K_{\mathrm{ext}}^{\mathrm{gm}}$,
the extrinsic (white-matter, wm) inputs $K_{\mathrm{ext}}^{\mathrm{wm}}$
and specific thalamic inputs $K_{\mathrm{ext}}^{\mathrm{th}}$. \prettyref{tab:External-inputs}
contains detailed information on the extracted values for all parameters
of external inputs.

\begin{table}
\begin{centering}
\begin{tabular}{rcccc}
\hline 
 & L2/3e  & L4e & L5e & L6e\tabularnewline
\hline
number of external clusters $\xi_{1}$ & 1 & 2 & 1 & 0\tabularnewline
number of external clusters $\xi_{2}$ & 2 & 1 & 0 & 1\tabularnewline
number of external clusters $\xi_{3}$ & 9 & 1 & 1 & 2\tabularnewline
number of clusters $\Xi_{\mathrm{total}}$ & 24 & 18 & 6 & 14\tabularnewline
cluster weight $\omega_{1}$ & 0.65 & 0.75 & 0.55 & 0.75\tabularnewline
cluster weight $\omega_{2}$ & 0.18 & 0.2 & 0.25 & 0.2\tabularnewline
cluster weight $\omega_{3}$ & 0.17 & 0.05 & 0.2 & 0.05\tabularnewline
number of local connections $k_{\mathrm{local}}$ & 4512 & 3277 & 2721 & 2702\tabularnewline
\hline 
number of gray-matter inputs $k_{\mathrm{ext}}^{\mathrm{gm}}$ & 534 & 353 & 389 & 79\tabularnewline
number of thalamic inputs & 0 & 93 & 0 & 47\tabularnewline
number of white-matter inputs & 1072 & 1665 & 1609 & 2790\tabularnewline
\hline 
total number of external inputs $k_{\mathrm{ext}}^{\mathrm{total}}$ & 1606 & 2111 & 1997 & 2915\tabularnewline
\hline
\end{tabular}
\par\end{centering}

\caption{Layer-specific external input properties. From top to bottom: the
extracted parameters required to calculate the number of gray-matter
external connections, the estimates of the three types of external
inputs (gray-matter, thalamic and white-matter) to excitatory neurons,
and the total number of external inputs. The total number of external
inputs is rounded for simulations (compare Table 4).\label{tab:External-inputs}}

\end{table}

\subsection*{Cell-type specific input structure}

To visualize the strongest pathways in the cell-type specific input
structure we calculate a input hierarchy. Therefore, we consider the
graph $G=(V,E)$ with the neuronal populations as vertices $V=\{\mathrm{L2/3e},\mathrm{L2/3i},\mathrm{L4e},\mathrm{L4i},\mathrm{L5e},\mathrm{L5i},\mathrm{L6e},\mathrm{L6i}\}$
and the edges $E=\{(v,w)|C((v,w))>0.04\}$ being those connections
with connection probabilities $C$ greater than $0.04$. In the input
hierarchy, the first order inputs to a vertex $w$ are defined as
the direct inputs\begin{eqnarray*}
L_{1}(w) & = & \{v|(v,w)\in E\}.\end{eqnarray*}
 The second order inputs to $w$ are the strongest of the direct inputs
to the elements in $L_{1}(w)$:

\begin{eqnarray*}
L_{2}(w) & = & \{v|\exists v\in V\setminus L_{1}(w)\\
 &  & \hspace{3pt}\hspace{3pt}\cap\exists w'\in L_{1}(w):(v,w')\in E\\
 &  & \hspace{3pt}\hspace{3pt}\cap\forall w''\in L_{1}(w)\setminus\{w'\}:C((v,w'))C((w',w))>C((v,w''))C((w'',w))\}.\end{eqnarray*}
The definition can easily be extended to higher order inputs. Second
and higher order inputs are excitatory (inhibitory) if the product
of mean synaptic strengths in the pathway is positive (negative).

\begin{figure}
\begin{centering}
\includegraphics{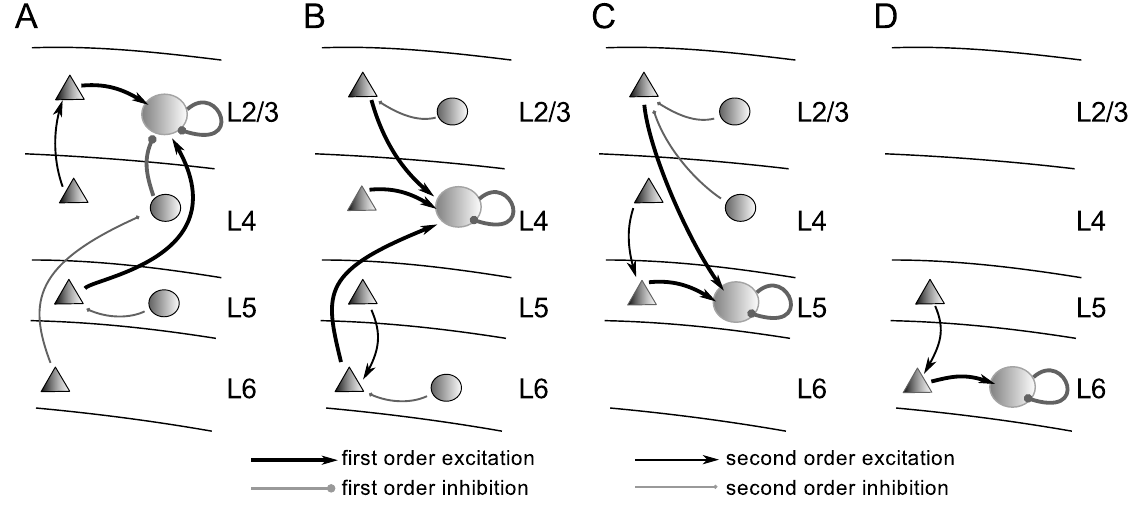}
\par\end{centering}

\caption{Differences in input structure for inhibitory cell types L2/3i (A),
L4i (B), L5i (C) and L6i (D) (large circles). The illustrations show
the strongest pathways of direct (first-order, thick arrows) and indirect
(second-order, thin arrows) excitation (black) and inhibition (gray)
of a given population. Triangles represent excitatory and circles
inhibitory populations.}

\end{figure}

\subsection*{Wiring algorithm}

\begin{figure}
\begin{centering}
\includegraphics{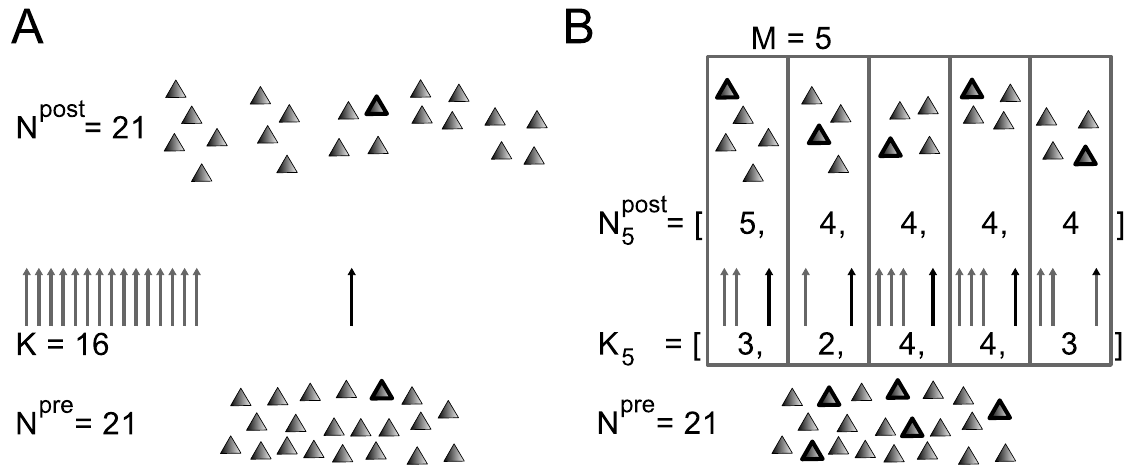}
\par\end{centering}

\caption{Wiring algorithm. The cartoons depicts the presynaptic (bottom triangles)
and the postsynaptic (top triangles) populations, here both containing
$N^{\mathrm{pre(post)}}=21$ neurons. The connection probability between
the two populations is $C=0.357$, corresponding to $K=16$ synapses
(arrows, black indicating the currently established synapses). (A)
Serial version. For every synapse, the pre- and postsynaptic neuron
(highlighted triangles) are drawn from the corresponding populations.
(B) Distributed version. The network is distributed over $M=5$ processes
(gray boxes) such that the first process hosts $\mathbf{N}_{5}^{\mathrm{post}}[0]=5$
postsynaptic neurons and all other processes $4$ postsynaptic neurons.
The number of synapses on each process (here $\textbf{K}_{M}=[3,2,4,4,3]$)
is multinomially distributed. The algorithm can establish $5$ connections
in parallel (black arrows and highlighted triangles).\label{fig:Wiring-algorithm}}

\end{figure}
The time consumption of the serial wiring algorithm described in \prettyref{fig:Wiring-algorithm}A
is $\mathcal{O}(K)$. In a distributed simulation setup (\prettyref{fig:Wiring-algorithm}B),
the neurons are distributed over $M$ processes and the synapses are
only created on the process where the postsynaptic neuron is located.
In this case, the high-level connection routine for randomly connecting
neuronal populations is as follows: Let $M$ be the number of processes,
$K$ the number of synapses to create, $\textbf{N}^{\mathrm{pre(post)}}$
the arrays containing all global identifiers (GIDs) of the presynaptic
(postsynaptic) neurons and $\textbf{N}_{M}^{\mathrm{post}}$ the distribution
of postsynaptic neurons on processes, i.e. $\textbf{N}_{M}^{\mathrm{post}}[m]$
contains all GIDs of postsynaptic neurons located on the $m$th process.
The numbers of synapses created on a given process $\textbf{K}_{M}$
is binomially distributed (because of the uniform probability that
a synapse is established on a given machine) with the boundary condition
that the total number of connections is $K=\sum_{m}\textbf{K}_{M}[m]$,
i.e. $\textbf{K}_{M}$ is multinomially distributed. In order to optimize
the parallelization of the high-level connection routine, we draw
a priori a random sample $\textbf{K}_{M}$ from the multinomial distribution
depending on $K$, $M$ and $\textbf{N}_{M}^{\mathrm{post}}$ \citep{Davis93_205}.
This sample has to be identical on every process, which can be achieved
by the usage of an identically seeded random number generator. To
create the synapses, we draw (on process $\mathrm{m}$ and using independent
random number generators on every process) $\textbf{K}_{M}[\mathrm{m}]$
times a presynaptic neuron $j$ from $\textbf{N}^{\mathrm{pre}}$
and a postsynaptic neuron $i$ from $\textbf{N}_{M}^{\mathrm{post}}[\mathrm{m}]$
as well as the synaptic weight $w_{ij}$ and delay $d_{ij}$ from
the gaussian distributions and call the low-level function \texttt{Connect($i$,$j$,$w_{ij}$,$d_{ij}$)}.
\prettyref{fig:Wiring-algorithm}B shows that the presynaptic neuron
is drawn from all neurons in the presynaptic population, but that
the postsynaptic neuron on a given process is drawn only from postsynaptic
neurons located on this process.

The algorithm requires a small serial overhead in order to create
the distribution of the postsynaptic neurons over processes and the
random sample from the multinomial distribution. Else, it guarantees
that on every process, only calculations are carried out that are
required to create synapses local to the process, i. e. the time consumption
is $\mathcal{O}(K/M)$.

\begin{figure}
\begin{centering}
\includegraphics{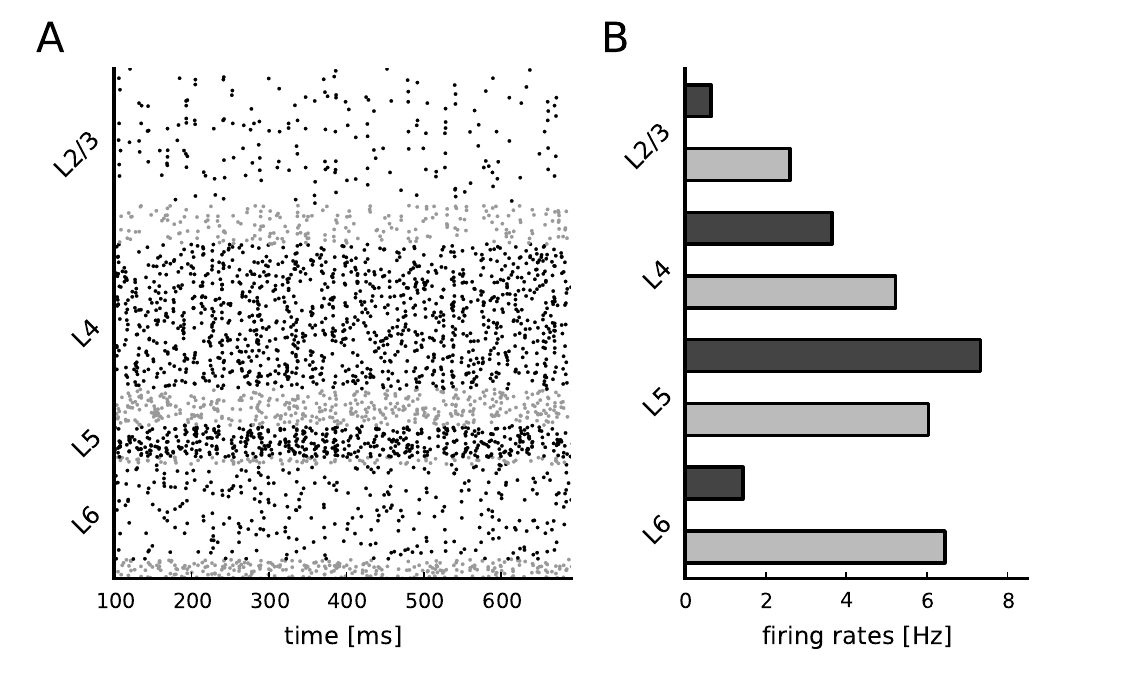}
\par\end{centering}

\caption{Replacing the Poissonian background noise with constant input currents
to all neurons yields basically the same activity, demonstrating that
the external noise is not requirement for the asynchronous irregular
firing. The raster plot (A) exhibits only slightly more synchrony
than the simulation with Poissonian background (Fig. 7A). The input
current to the different populations is chosen identical to the mean
input current evoked by the Poissonian background input in the reference
parametrization (Table 4). The resulting firing rates (B) are the
same as in Fig. 7B.}

\end{figure}

\begin{figure}
\begin{centering}
\includegraphics{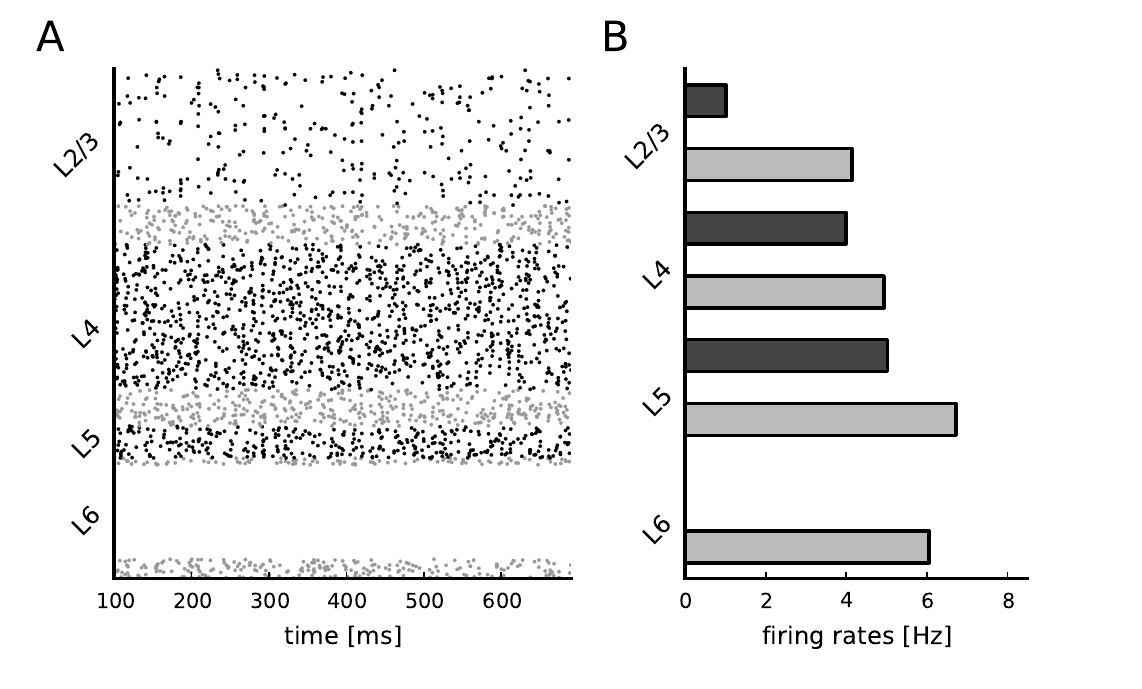}
\par\end{centering}

\caption{The network simulations exhibit comparable spontanous activity features
when the external inputs are independent of the layer. (A) shows the
raster plot and (B) the cell-type specific firing rates after replacing
the estimated numbers of external inputs $k_{\mathrm{ext}}$ with
$k_{\mathrm{ext}}^{\mathrm{hom}}=2000$ for excitatory populations
and $k_{\mathrm{ext}}^{\mathrm{hom}}=1850$ for inhibitory populations.
The order of firing rates is basically the same as with data based
external inputs: L2/3e and L6e exhibit lowest, L4e intermediate and
L5e highest rates; inhibitory rates are higher than excitatory ones.
L6e is basically silent. This is due to the small excitatory input
to the excitatory population: the data based estimate assigns 50\%
more inputs to this population.}

\end{figure}

\begin{figure}
\begin{centering}
\includegraphics{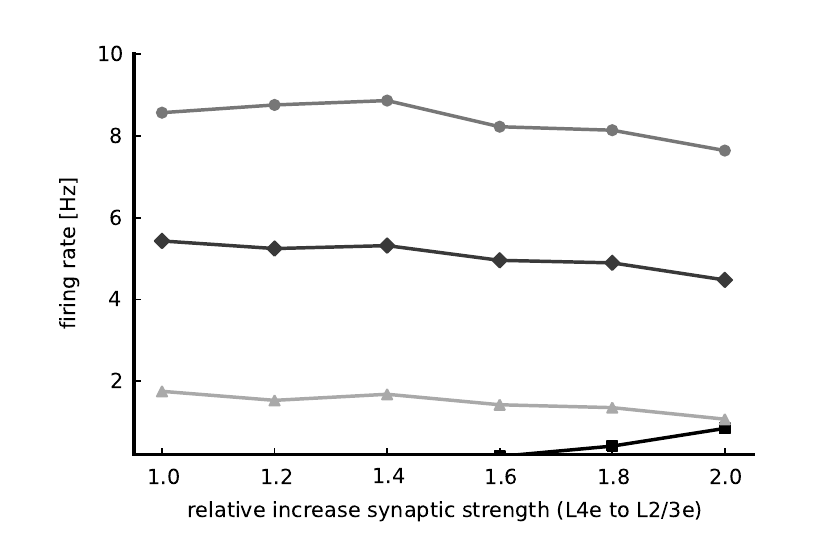}
\par\end{centering}

\caption{Dependence of stationary activity on the synaptic strength of the
L4e to L2/3e connection. The available experimental data do not constrain
the connection probabilities between L4 and L2/3 unambigously (see
also \emph{Discussion}). \citet{Feldmeyer06_583} find that the convergence
onto L2/3 pyramids from other L2/3 pyramids and from L4 cells is the
same. In our model, however, the convergence from L2/3 cells is twice
as large as from L4. To test for the influence of this connection,
we increase the synaptic strength of the L4e to L2/3e connection.
The graph shows the firing rates of the excitatory populations in
L2/3 (squares), L4 (diamonds), L5 (circles) and L6 (triangles), lightness
increases with depth as a function of the relative increase of this
synaptic strength. L2/3e fires a significant number of spikes for
an increase of $>1.6$. The other layers are only slightly affected
by this modification. An increase of $2$ is consistent with the relative
convergence described by \citet{Feldmeyer06_583} and corresonds to
our reference parametrization.}

\end{figure}

\begin{figure}
\begin{centering}
\includegraphics{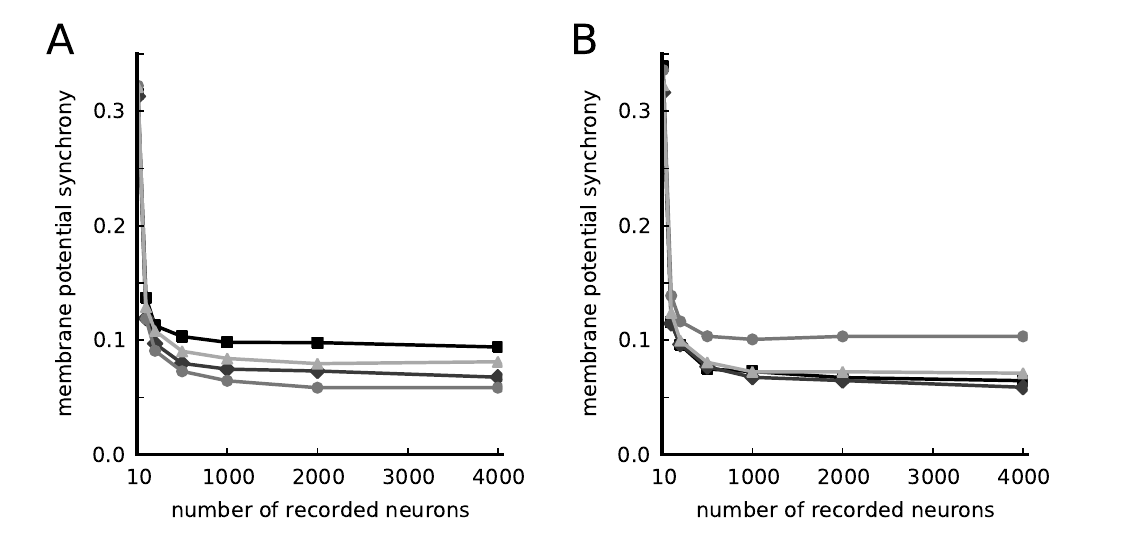}
\par\end{centering}

\caption{Membrane potential synchrony according to \citep{Golomb07_1347}.
The graphs shows the membrane potential synchrony of (A) excitatory
populations and (B) inhibitory populations in L2/3 (squares), L4 (diamonds),
L5 (circles) and L6 (triangles), lightness increases with depth, as
a function of the number of recorded membrane potential traces. The
membrane potential synchrony $\chi$ depends on the number of neurons
$N$ as $\chi(N)=\chi(\infty)+\frac{a}{\sqrt{N}}+\mathcal{O}(N)$
where $a>0$ is a constant \citep{Golomb07_1347}. As known from previous
studies on mono-layer models \citep{Brunel00}, the asynchronous activity
is not perfectly asynchronous (which would correspond to $\chi(\infty)=0$),
but weakly synchronized. In the layered network model, the subthreshold
activity is similar, with $\chi(\infty)\approx0.1$. The membrane
potential synchrony of excitatory neurons is ordered as follows (from
largest to smallest values): L2/3e, L6e, L4e, L5e. For inhibitory
neurons it is highest in L5i and almost identical in the other layers.}

\end{figure}

\newpage

\begin{longtable}{l>{\raggedright}p{1.8cm}>{\raggedright}p{1.8cm}>{\raggedright}p{1.6cm}>{\raggedright}p{1.6cm}>{\raggedright}p{1.8cm}>{\raggedright}p{1.8cm}}
\caption{Numerical connectivity values. From left to right: connection specifier,
physiological connection probability estimates for raw and modified
data, anatomical estimates for the number of synapses (in billion)
and the corresponding connection probabilities for raw and modified
data sets, respectively. A dash (--) indicates connections that have
not been measured. Non-zero but undetermined values of the raw physiological
data set are given with the estimates used by \citet{Haeusler07_149}
in brackets. \label{tab:Numerical-C-values}}
\tabularnewline
\hline 
connection & $C_{\mathrm{p}}^{\mathrm{raw}}$ & $C_{\mathrm{p}}^{\mathrm{mod}}$ & $K_{\mathrm{a}}^{\mathrm{raw}}$ & $K_{\mathrm{a}}^{\mathrm{mod}}$ & $\tilde{C}_{\mathrm{a}}^{\mathrm{raw}}$ & $\tilde{C}_{\mathrm{a}}^{\mathrm{mod}}$\tabularnewline
\hline
\hline
\endfirsthead
\hline 
connection & $C_{\mathrm{p}}^{\mathrm{raw}}$ & $C_{\mathrm{p}}^{\mathrm{mod}}$ & $K_{\mathrm{a}}^{\mathrm{raw}}$ & $K_{\mathrm{a}}^{\mathrm{mod}}$ & $\tilde{C}_{\mathrm{a}}^{\mathrm{raw}}$ & $\tilde{C}_{\mathrm{a}}^{\mathrm{mod}}$\tabularnewline
\hline
\hline
\endhead
L2/3e$\to$L2/3e & 0.223 & 0.122 & 73.6 & 73.6 & 0.172 & 0.172\tabularnewline
L2/3e$\to$L2/3i & 0.212 & 0.416 & 10.3 & 10.3 & 0.086 & 0.086\tabularnewline
L2/3i$\to$L2/3e & 0.182 & 0.413 & 16.7 & 20.3 & 0.139 & 0.169\tabularnewline
L2/3i$\to$L2/3i & 0.4 & 0.4 & 2.9 & 3.4 & 0.086 & 0.101\tabularnewline
\hline 
L4e$\to$L4e & 0.174 & 0.122 & 23.5 & 23.5 & 0.05 & 0.05\tabularnewline
L4e$\to$L4i & 0.19 & 0.282 & 3.5 & 3.5 & 0.029 & 0.029\tabularnewline
L4i$\to$L4e & 0.095 & 0.304 & 10.8 & 18.0 & 0.09 & 0.15\tabularnewline
L4i$\to$L4i & 0.5 & 0.5 & 1.9 & 2.9 & 0.065 & 0.098\tabularnewline
\hline 
L5e$\to$L5e & 0.092 & 0.122 & 3.1 & 3.1 & 0.13 & 0.13\tabularnewline
L5e$\to$L5i & 0.096 & 0.109 & 0.42 & 0.42 & 0.081 & 0.081\tabularnewline
L5i$\to$L5e & 0.123 & 0.194 & 0.42 & 4.0 & 0.082 & 0.781\tabularnewline
L5i$\to$L5i & 0.6 & 0.6 & 0.07 & 0.47 & 0.062 & 0.414\tabularnewline
\hline 
L6e$\to$L6e & -- & -- & 15.7 & 15.7 & 0.075 & 0.075\tabularnewline
L6e$\to$L6i & -- & -- & 1.4 & 1.4 & 0.032 & 0.032\tabularnewline
L6i$\to$L6e & -- & -- & -- & 11.4 & -- & 0.269\tabularnewline
L6i$\to$L6i & -- & -- & -- & 1.5 & -- & 0.173\tabularnewline
\hline 
L2/3e$\to$L4e & 0.0 & 0.0 & 9.2 & 9.2 & 0.02 & 0.02\tabularnewline
L2/3e$\to$L4i & 0.163 & 0.163 & 0.95 & 0.95 & 0.008 & 0.008\tabularnewline
L2/3i$\to$L4e & 0.0 & 0.0 & 1.8 & 1.8 & 0.014 & 0.014\tabularnewline
L2/3i$\to$L4i & 0.0 & 0.0 & 0.22 & 0.22 & 0.007 & 0.007\tabularnewline
\hline 
L2/3e$\to$L5e & 0.58 & 0.180 & 12.3 & 12.3 & 0.123 & 0.123\tabularnewline
L2/3e$\to$L5i & 0.0 & 0.0 & 1.5 & 1.5 & 0.067 & 0.067\tabularnewline
L2/3i$\to$L5e & $>$0.0(0.2) & $>$0.0(0.2) & 0.84 & 0.84 & 0.03 & 0.03\tabularnewline
L2/3i$\to$L5i & $>$0.0(0.0) & $>$0.0(0.0) & 0.08 & 0.08 & 0.013 & 0.013\tabularnewline
\hline 
L2/3e$\to$L6e & -- & -- & 9.2 & 9.2 & 0.031 & 0.031\tabularnewline
L2/3e$\to$L6i & -- & -- & 0.26 & 0.26 & 0.004 & 0.004\tabularnewline
L2/3i$\to$L6e & -- & -- & 0.66 & 0.66 & 0.008 & 0.008\tabularnewline
L2/3i$\to$L6i & -- & -- & 0.02 & 0.02 & 0.001 & 0.001\tabularnewline
\hline 
L4e$\to$L2/3e & 0.147 & 0.112 & 18.4 & 18.4 & 0.041 & 0.041\tabularnewline
L4e$\to$L2/3i & 0.098 & 0.098 & 2.6 & 2.6 & 0.02 & 0.02\tabularnewline
L4i$\to$L2/3e & 0.327 & 0.327 & 0.96 & 0.96 & 0.008 & 0.008\tabularnewline
L4i$\to$L2/3i & $>$0.0(0.2) & $>$0.0(0.2) & 0.17 & 0.17 & 0.005 & 0.005\tabularnewline
\hline 
L4e$\to$L5e & 0.0 & 0.14 & 3.7 & 3.7 & 0.035 & 0.035\tabularnewline
L4e$\to$L5i & 0.0 & 0.0 & 0.42 & 0.42 & 0.018 & 0.018\tabularnewline
L4i$\to$L5e & 0.0 & 0.0 & 0.36 & 0.36 & 0.014 & 0.014\tabularnewline
L4i$\to$L5i & 0.0 & 0.0 & 0.03 & 0.03 & 0.005 & 0.005\tabularnewline
\hline 
L4e$\to$L6e & -- & -- & 7.8 & 7.8 & 0.025 & 0.025\tabularnewline
L4e$\to$L6i & -- & -- & 0.26 & 0.26 & 0.004 & 0.004\tabularnewline
L4i$\to$L6e & -- & -- & 1.6 & 1.6 & 0.02 & 0.02\tabularnewline
L4i$\to$L6i & -- & -- & 0.01 & 0.01 & 0.001 & 0.001\tabularnewline
\hline 
L5e$\to$L2/3e & 0.034 & 0.014 & 8.9 & 8.9 & 0.089 & 0.089\tabularnewline
L5e$\to$L2/3i & 0.0 & 0.0 & 1.3 & 1.3 & 0.045 & 0.045\tabularnewline
L5i$\to$L2/3e & 0.0 & 0.0 & 0.0 & 0.0 & 0.0 & 0.0\tabularnewline
L5i$\to$L2/3i & 0.0 & 0.0 & 0.0 & 0.0 & 0.0 & 0.0\tabularnewline
\hline 
L5e$\to$L4e & 0.0 & 0.0 & 1.7 & 1.7 & 0.016 & 0.016\tabularnewline
L5e$\to$L4i & 0.0 & 0.0 & 0.21 & 0.21 & 0.008 & 0.008\tabularnewline
L5i$\to$L4e & 0.0 & 0.0 & 0.18 & 0.18 & 0.001 & 0.001\tabularnewline
L5i$\to$L4i & 0.0 & 0.0 & 0.0 & 0.0 & 0.0 & 0.0\tabularnewline
\hline 
L5e$\to$L6e & -- & -- & 4.8 & 4.8 & 0.068 & 0.068\tabularnewline
L5e$\to$L6i & -- & -- & 0.47 & 0.47 & 0.033 & 0.033\tabularnewline
L5i$\to$L6e & -- & -- & 0.36 & 0.36 & 0.024 & 0.024\tabularnewline
L5i$\to$L6i & -- & -- & 0.03 & 0.03 & 0.01 & 0.01\tabularnewline
\hline 
L6e$\to$L2/3e & -- & -- & 2.7 & 2.7 & 0.009 & 0.009\tabularnewline
L6e$\to$L2/3i & -- & -- & 0.42 & 0.42 & 0.005 & 0.005\tabularnewline
L6i$\to$L2/3e & -- & -- & -- & 0.0 & -- & 0.0\tabularnewline
L6i$\to$L2/3i & -- & -- & -- & 0.0 & -- & 0.0\tabularnewline
\hline 
L6e$\to$L4e & -- & -- & 37.2 & 37.2 & 0.12 & 0.12\tabularnewline
L6e$\to$L4i & -- & -- & 5.7 & 5.7 & 0.072 & 0.072\tabularnewline
L6i$\to$L4e & -- & -- & -- & 0.0 & -- & 0.0\tabularnewline
L6i$\to$L4i & -- & -- & -- & 0.0 & -- & 0.0\tabularnewline
\hline 
L6e$\to$L5e & -- & -- & 1.7 & 1.7 & 0.024 & 0.024\tabularnewline
L6e$\to$L5i & -- & -- & 0.16 & 0.16 & 0.01 & 0.01\tabularnewline
L6i$\to$L5e & -- & -- & -- & 0.0 & -- & 0.0\tabularnewline
L6i$\to$L5i & -- & -- & -- & 0.0 & -- & 0.0\tabularnewline
\hline
mean values & 0.145 & 0.139 & -- & -- & 0.046 & 0.079\tabularnewline
\hline
\end{longtable}
\end{document}